\documentclass[
aps,nofootinbib,
showpacs,showkeys,preprint
tightenlines,preprintnumbers,] {revtex4}

\usepackage{epsf,epsfig,subfigure,graphicx,amsmath,amssymb 
}
\usepackage{color}
\newcommand{\dis}[1]{\begin{equation}\begin{split}#1\end{split}\end{equation}}
\newcommand{\ie}{{\it i.e.~}}
\newcommand{\etal}{{\it et al.\,}} 

\newcommand{\QPQ}{Q_{\rm PQ}}
\newcommand{\QDE}{Q_{\rm DE}}

\newcommand{\tev}{\,\textrm{TeV}}
\newcommand{\gev}{\,\textrm{GeV}}
\newcommand{\meV}{\,\mathrm{MeV}}

\newcommand{\eV}{\,\mathrm{eV}}
\newcommand{\Mp}{M_{\rm P}}

\newcommand{\Mg}{{M_{\rm GUT}}}

\newcommand{\NDW}{N_{\rm DW}} 

\newcommand{\Lqcd}{\Lambda_{\rm QCD}}

\newcommand{\USW}{U(1)$_{\rm global}\times \rm SU(2)_W^2$}
\newcommand{\Uanom}{U(1)$_{\rm anom}$}
\newcommand{\alem}{\alpha_{\rm em}}
\newcommand{\Qem}{Q_{\rm em}}

\newcommand{\UoR}{U(1)$_{\rm R}$}
\newcommand{\UDE}{U(1)$_{\rm DE}$}  
  
\newcommand{\UPQ}{U(1)$_{\rm PQ}$}

\def\sw0{{$\sin^2\theta_W^0$}}

\def\Tr{{\rm Tr \,}}

\def\Nf2{{\bf N_{[2]}}}

\newcommand{\Z}{{\bf Z}}
\newcommand{\ZR}{{\bf Z}_{2R}}
\newcommand{\ZZR}{{\bf Z}_{4R}}

\def\SU2Ch{SU(2)$_L\times$SU(2)$_R$}
\def\E6{{\rm E_6}}

\def\EE8{{\rm E_8\times E_8'}}

\def\flip{SU$(5)_{\rm flip}$}

\def\one{{\bf 1}}

\def\five{{\bf 5}}
\def\ten{{\bf 10}}
\def\tenb{{\overline{\bf 10}}}

\def\fiveb{{\overline{\bf 5}}}

\def\nine{{\bf 9}}
\def\nineb{\overline{\bf 9}\,}

\def\tsixb{\overline{\bf 36}\,}
\def\tsix{{\bf 36}}

\def\one{\bf 1}

\def\five{\bf 5}
\def\ten{\bf 10}
\def\tenb{\overline{\bf 10}} 

\begin{document}

\draft

\title{\Large\bf Quintessential Axions
}

\author{Jihn E.  Kim}
\address{Department of Physics, Seoul National University, 1 Gwanak-Ro, Seoul  08826, Republic of Korea,  
}
\address{Department of Physics, Kyung Hee University, 26 Gyungheedaero, Dongdaemun-Gu, Seoul 02447, Republic of Korea  
}
 
\begin{abstract} 
We  
 review the dark energy solutions by a very light pseudoscalar called ``quintessential axion''.  
For the explicit breaking terms, we consider both the global anomaly \USW~and the potential $\Delta V$.  At the field theory level, we will present a strategy for picking up one such pseudoscalar among plenty of pseudoscalars. At string level, numerous antisymmetric tensor fields are pseudoscalars. Including these, we present  a confining force example  via a $\Z_{12-I}$ orbifold compactification of SO(32) heterotic string.  In addition, we also aim to draw attention on almost massless mesons of confining nonabelian gauge group, which can be another motivation for introducing an additional confining force,  to explain dark energy as the vacuum energy of these mesons. 

\end{abstract}
\maketitle

\section{Introduction}\label{sec:Introduction}

Light pseudoscalar particles are the lampposts to the road leading to physics scales much above their masses. The well-known example is the pion triplet which guided toward the \SU2Ch~chiral symmetry glanced at the strong interaction scale \cite{Nambu61,GellMann68}. Another is the very light axion which hints the intermediate scale \cite{KimPRL79,KNS18}. The scale where the symmetry is explicit is
$E\gtrsim f\sim\Lambda^2/m$, for the pseudoscalar mass $m$, where $\Lambda^4$ is a typical energy density contributed by such a pseudoscalar. Before obtaining mass by the energy density  perturbation of order $\Lambda^4$, these pseudoscalar degrees are phase degrees $\theta(x)$ of some unitary operators such that they have kinetic energy terms in quantum field theory below the defining scale $f$ of the  pseudoscalar. Above the scale $f$, the phase is not a dynamical field, \ie not depending on $x$, but it still represents a phase direction of the global symmetry, but gravity breaks this global symmetry even at the Planck scale.  

If the pseudoscalar mass $m$ is less than 1 MeV, it can decay to two photons and two neutrinos. For these two rare processes, the $\pi^0$ decays are predicted to be  $\Gamma(\pi^0\to 2\gamma)\simeq ( \alem^2/64\pi^3) (M_{\pi^0}/f_\pi)^2M_{\pi^0}$ (at leading order \cite{Holstein02}) and $1.2\times 10^{-5}\Gamma(\pi^0\to 2\gamma)$ \cite{Shrock79}, respectively. For a pseudoscalar to be still present in our Universe, therefore we require $\Gamma^{-1} > 4.3\times 10^{17}{\rm s} = 1/(1.53\times 10^{-33} \eV)$.
 Using the above $f-m$ relation, we get an idea on $m$ from the condition that it survives until now. If $\Lambda$ is the QCD scale 
 \dis{
 \Gamma^{-1}\simeq  \frac{ 64\pi^3\,\Lambda_{\rm QCD}^4}{\alem^2m^4}   \frac{1}{m}>\frac{1}{1.53\times 10^{-33} \eV}
 }
or $m< 65\, \eV$ for $\Lambda_{\rm QCD}=380\,\meV$.\footnote{For the very light axion, the specific coupling reduces it further to 24\,eV \cite{KimRMP10}.} If we take $\Lambda^4$ as the current energy density of the Universe, $(0.003\,\eV)^4$, then pseudoscalars with mass less than $3.3\times 10^{-4}\, \eV$ survives until now. Therefore, very light pseudoscalars attracted a great deal of attention in cosmology. Pseudoscalars for  $m<3.3\times 10^{-4}\, \eV$ can contribute to the current energy density and structure formation. In this context, quinessential axion (QA) \cite{Carroll98,Hill02,KN03} was suggested for dark energy (DE), and ultra-light axion (ULA) \cite{KimPRD16} was suggested for the galactic scale structures \cite{Witten17}. The precursers of these pseudoscalars are `general quintessence', not pinpointing to the pseudoscalars \cite{quinessence}.  

Not only to DE, but also the axion contributtion to dark matter (DM) in the universe is possible with QCD axion \cite{PWW83,AS83,DF83,Bae08},
\dis{ 
\frac{\Omega_a^{\rm QCD}}{0.2}\approx (0.391\times 0.701/h)^2\left(\frac{10^{12}\gev}{f_a}\right)^{1.184}F(\theta_i)\theta_i^2,~\textrm{for }\Lambda_{\rm QCD}=380\meV  
}
where $f_a$ is the decay constant of the QCD axion, $\theta_i$ is the initial misalignment angle, $F(\theta_i)$ is O(1) function \cite{Bae08}, $\Lambda_{\rm QCD}$ is the dimensional transmutation scale of QCD, and $h$ is the Hubble constant today in units of 100 km/s/Mpc. We will consider this QCD axion also in this paper because it is present in most models.
Note that quintessential axion mass near $10^{-32\,}\eV$ is constrained from the evolution history of the universe \cite{ChoiK21,LNP954,Marsh21}, but there remains a room for DE to be accomodated. Cosmological evolution of axions produces domain walls  in a spontaneously broken gauge theory \cite{Kibble77}, which is problematic if the domain-wall number, $\NDW$, is not 1.  Note, however, that the  $\NDW= 1$ models can be easily accomodated \cite{Sikivie82DW,Barr87}.

In string theory, there are antisymmetric tensor fields which are pseudoscalars \cite{Witten84,Witten85,ChoiK85A,ChoiK85B}, and hence $f$ in this case is the string scale $M_s$. Below the compactification scale, various scales combine to give $f$ above \cite{KNP05} or below \cite{KimPLB17} the string scale, depending on the compactification schemes. In this paper, however, we will not discuss this kind of ultraviolet completed theories because they depend on the details at the Planck scale physics \cite{ConlonSven16}.

In Sec.  \ref{sec:QAs}, we present the QA models, and in Sec. \ref{sec:Axions} possible axionic models are discussed, including $\Delta V$ for the quitessential axion.  In Sec. \ref{sec:Rparity}, R-parity in supersymmetric models are presented, and in Sec. \ref{sec:String} a $\Z_{12-I}$ orbifold compactification of SO(32) hetrerotic string toward QA is reviewed. Section   \ref{sec:Conclusion} is a brief conclusion. 

\section{Quintessential Axions}\label{sec:QAs}

For a pseudoscalar created by spontaneous breaking of a global symmetry, with the phase represented by  $e^{i\theta}$, there is no pseudoscalar above the defining scale $f$. Below $f$, the fundamental scale for the pseudoscalar decay-interactions is $f$ and hence  $f$  is called the {\it decay constant}. The global transformation is expressed as  $\theta\to a/f$, \ie $e^{i\,a/f}$ where $a$ is the pseudoscalar.
On the other hand, if a pseudoscalar is present in the theory, then $f$ is the defining scale of that theory. In this sense, string scale $M_s$ itself is the decay constant for axions from string theory.  

In the effective field theory approach, pseudoscalars are discussed in relation to some global symmetry. Here, some  important scales are given above or below  the global symmetry breaking scale $f$.  Such a scale above $f$ is the string/Planck scale or the GUT scale.  Such a scale below $f$ is some confining scale   of a new confining force if it is present.  Another scale we need is the explicit symmetry breaking scale $\Lambda$. One class of  $\Lambda$ is in the potential $V$ of spin-0 fields and Yukawa couplings, \ie via interactions, and the second class is the non-abelian gauge anomalies.  Since gravitational interactions always break any global symmetry \cite{BanksDine92},  we can consider the explicit breaking term at $\Lambda$ due to gravity also, of order $\Lambda^{n+4}/\Mp^n$. For this term to be sufficiently small for QA or ULA, theory must possess some symmetry to forbid gravity-originated terms up to the order $\Lambda^{n+3}/\Mp^{n-1}$.
Of course, the presence of some operators but not of others is rather artificial and in addition unnatural unless there is some underlying symmetry which justifies it.\footnote{In Ref.  \cite{KimPRD16}, discrete symmetries were used to realize this scheme.} 
In this paper we review  the key ideas of QA.  

\subsection{Mesons from confining force}
Note that pions appear  below the condensation scale of  QCD quarks. Similarly, if there is a confining force whose scale is near the string scale, mesons realized by the condensation of ex-quarks of the extra confining force can be  candidates for QA, which has been proposed recently \cite{KimKimNam22}. 

Suppose that the flavor symmetry of ex-quarks is U$(N)\times$U$(N)$ which reduces to SU$(N)\times$SU$(N)$ due to the anomaly breaking of U$(N)\times$U$(N)$.   The light mesons belonging to  SU$(N)\times$SU$(N)$ obtain mass by explicit breaking of the flavor symmetry of  ex-quarks. The explicit breaking is through gravity where the breaking terms are suppressed by the Planck mass $\Mp$. To present an example, we need a detail model. 

Later we will present a general idea along this line at field theory level and show a realization in string compactification. The explicit breaking terms are given at the string/Planck scale.

\subsection{Goldstone bosons from global symmetry}

A Goldstone boson can be related to a global symmetry, for which let us introduce  \UDE. With   \UDE, the essential component is a SM singlet complex scalar field $\sigma$ \cite{KimPRL79}. The decay constant $f$ is the vacuum expectation value (VEV) of $\sigma$.    The explicit breaking terms can appear as non-Abelian gauge anomalies, and as terms in the potential $V$ and/or in the Yukawa couplings ${\cal L}_Y$. The Peccei-Quinn (PQ) symmetry \cite{PQ77} is assumed to be broken by the QCD anomaly to provide DM of the universe. So, the QCD anomaly   will be  separated from the DE part. The U(1) charges $Q_{\rm PQ}$ and  $Q_{\rm DE}$ of the singlet field  $\sigma$ and SM fields are assumed to be related by terms
\dis{
&Q_{\rm PQ}:~(\sigma^*)(H_uH_d)^n,\\
&Q_{\rm DE}:~(\sigma^*)^{m'}(H_uH_d)
}  
\begin{table}[h!]
\begin{center}
\begin{tabular}{@{}|c| c c c|cccccc|@{}} \hline
 &$H_u,$&  $H_d,$ &    $\sigma$& $q_L$& $t_R$& $b_R$& $\ell_L$& $e_R$& $N_R$\\[0.2em] \colrule &&&&  \\[-1.4em] 
 ~$Q_{\rm PQ}$~  & $1$  & $1$ &   $2n$ &$\frac{1}2$&$-\frac{1}2$&$-\frac{1}2$ &$\frac{1}2$&$-\frac{1}2$&$-\frac{1}2$\\[0.3em]  \\[-1.4em] 
~$Q_{\rm DE}$~  & $\frac{m'}{2}$  & $\frac{m'}{2}$ &   $1$ &$\frac{m'}{4}$  &$\frac{-m'}{4}$  &$\frac{-m'}{4}$  &$\frac{m'}{4}$&$\frac{-m'}{4}$ &$\frac{-m'}{4}$ \\[0.2em]
 \hline
\end{tabular}
\end{center}
\caption{The   \UPQ~and \UDE~charges. $Q_{\rm PQ}(Q_{\rm DE})$ of $H_{u,d}(\sigma)$ is defined as 1.}\label{tabQs}
\end{table}
If the Yukawa couplings preserve the \UPQ~and \UDE~symmetries, the global charges  are as given in Table \ref{tabQs}. A QCD axion resulting from  Table \ref{tabQs} with an additional heavy quark $Q$ is the heavy quark axion model \cite{KimPRL79, Shifman80}.

\subsubsection{Anomaly breaking}\label{subsec:Anomaly}

Non-abelian gauge forces break global symmetries \cite{Belavin75}. The breaking term at the scale $\mu$ is proportional to
\dis{
e^{-2\pi/\alpha_a(\mu)}
}
where $\alpha_a(\mu)$ is the gauge coupling of the non-Abelian gauge group $G_a$. If $\alpha_a$ is small as in the weak gauge group SU(2)$_W$ in the SM, the coefficient is extremely small. However, for strong interaction  SU(3)$_c$ in the SM the coefficient can be of order 1 where  $\alpha_c$ becomes order 1.

\subsubsection{Breaking of \UDE\,by terms in the potential}\label{subsec:Vs}

For the singlet $\sigma$, any power of $\sigma^*\sigma$ preserves the \UDE\,symmetry. In terms of $\sigma$, the explicit breaking terms take the form $(\sigma^*\sigma)^i\sigma^j+{\rm h.c.}\,(i,j=\rm integers)$. If all terms up to a large $2i+j$ are excluded from some discrete symmetry as done in  \cite{KimPRD16}, the breaking can be made very tiny.  Standard model singlets composed of the SM scalar fields are powers of $H_u H_d$. So, the explicit breaking terms available below a scale $M$ take the forms
\dis{
&\frac{1}{(M)^{2i+j+2l-4}}(\sigma^*\sigma)^i\sigma^j (H_u H_d)^l+{\rm h.c.}\\
&\frac{1}{(M)^{2i'+j'+2l'-4}}(\sigma^*\sigma)^{i'}\sigma^{*j'} (H_u H_d)^{l'}+{\rm h.c.}\label{eq:ExBreakingV}
}
where $i,j,l,i',j',l'=\rm integers$, and  inverse powers of the compactification scale mass $M$ are shown.
 
\subsubsection{Breaking of \UDE\,by Yukawa couplings}\label{subsec:Yukawas}

Since fermions are the SM quarks and leptons, the explicit breaking terms  take the following forms of Yukawa couplings  
\dis{
(1+\delta_1)\frac{f_d}{\sqrt2} \overline{q}_L d_R H_d,\\
(1+\delta_2)\frac{f_u}{\sqrt2} \overline{q}_L d_R H_u,\\
(1+\delta_e)\frac{f_e}{\sqrt2} \overline{\ell}_L e_R H_d,\label{eq:ExBreakingY}
}
where 1's correspond to the SM Yukawa couplings, and $\delta$'s represent the degree of explicit breaking of the global symmetry.  $\delta$'s can be replaced by the SM singlets breaking the global symmetries as shown in Eq. (\ref{eq:ExBreakingV}).

\subsubsection{Gravity effects}\label{subsec:GravCont}

Gravity effects of breaking global symmetries can be given by the terms explicitly breaking the global symmetries.  Firstly, in the potential $\Delta V=m^3\sigma +{\rm h.c.}$ breaks the global symmetry where $m$ is considered to be the Planck scale. For the QCD axion, this effect was noticed by several groups \cite{BarrSeckel92,Kamionkowski92,Holman92}.
Second, in the anomaly breaking of  SU(2) gauge
theory one might argue that the SU(2) vacuum angle $\theta_{\rm SU(2)}$ 
can be rotated away by anomalies for the global
$B + L$ symmetry, where $B$ is the baryon number and $L$ is the lepton number. But, the shift of $\theta_{\rm SU(2)}$ is not a flat direction due to  gravity effects of breaking global symmetries. So, the shift of $\theta_{\rm SU(2)}$ is considered physical.

\section{Axions}\label{sec:Axions}
 
All pseudoscalars can be expressed as phase fields, $e^{i\theta_i T^i}$ where $T^i$ is the group generator and $\theta=a/f$. As mentioned before, $f$ is the defining scale of the pseudoscalar. Customarily, it is equivalent to an effective VEV breaking the corresponding global symmetry. Among these pseudoscalars, if the leading breaking term is an   anomalous term of (global symmetry)--(non-Abelian gauge group)--(non-Abelian gauge group), the pseudoscalar is called an axion. The first example was given from the anomaly \UPQ-SU(3)$_{\rm color}$-SU(3)$_{\rm color}$ \cite{PQ77}.

\subsection{QCD axion}\label{subsec:QCDaxion}

\begin{figure}[!t]\hskip -0.3cm
\includegraphics[height=0.45\textwidth]{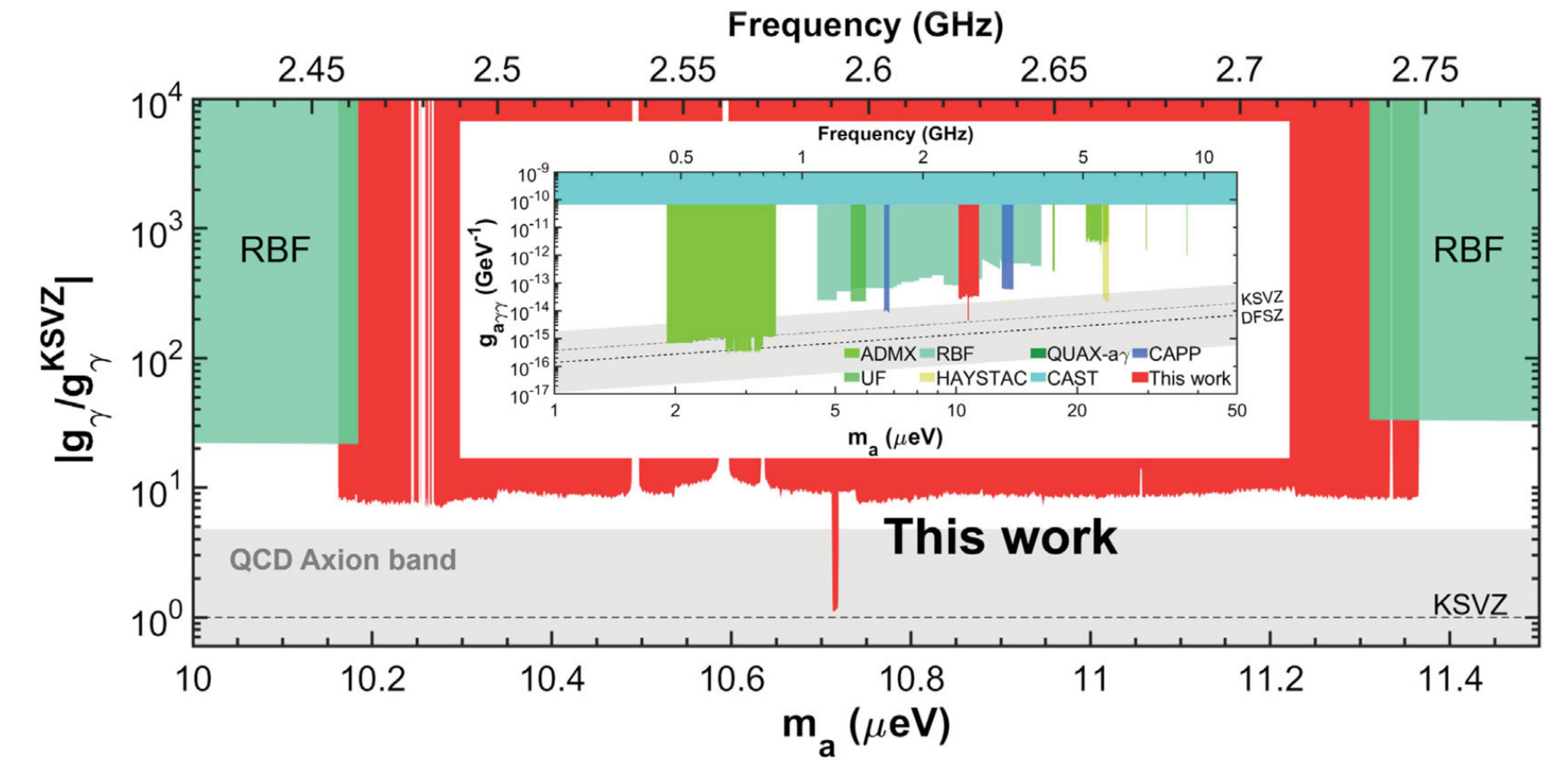}
\caption{\label{fig:QCDaxion}  The CAPP-PACE exclusion limit at 90\% confidence level (red area) \cite{CAPPprl21}. Vacancy between the Rochester-Brookhaven-Fermilab
(RBF) results (mint color gamut) [14] is filled with this work. The inset shows this work along with other axion searching results in the
extended axion mass range }
\end{figure}

\UPQ, having color anomaly, asks for  assigning charges to some colored fermions, notably to quarks. In the SM, only spin-$\frac12$ quarks carry color. With a supersymmetric extension, spin-$\frac12$ gluinos also carry color. Using only the SM quark doublets, $\Qem=+\frac23\,(\ie T_3=+\frac12)$ unit up-type quarks and  $\Qem=-\frac13\,(\ie T_3=-\frac12)$ unit down-type quarks are assigned with different PQ charges $\QPQ$ \cite{PQ77}, otherwise there is no color anomaly because $\Tr {T_3}=0$. This example was the first QCD axion, but soon it was ruled out from the results of beam dump experiments \cite{KimRMP10}. For an acceptable QCD axion, BSM quarks \cite{KimPRL79, Shifman80} should be introduced for the color anomaly, but the axion should reside in the phase of a BSM scalar $\sigma$ \cite{KimPRL79}, which was named as the Kim-Shifman-Vainshtein-Zakharov(KSVZ) axion \cite{Kimprp86}. For the so-called Dine-Fischler-Sredicki-Zhitnitzky(DFSZ) axion \cite{DFSa,Zhit}, only the SM quarks are used for colored fermions but with   interactions relating the PQ charges of $\sigma$ and the SM quarks. Nevertheless, a BSM scalar $\sigma$ is needed anyway to house the QCD axion as emphasized in Ref. \cite{KimPRL79}.  In the supersymmetric extension, gluinos are not enough for a QCD axion since the chief role of gluino condensation is to break (global) supersymmetry. Here again a BSM scalar $\sigma$ is needed to house a QCD axion. A recent compilation of the allowed parameter regions of the QCD axion will be  shown later in Fig. \ref{fig:QCDaxion}.

In the cosmic evolution, discrete symmetries belonging to the \UPQ~is better to be a subgroup of a gauge group not to lead to some kind of domain wall problem. This kind of discrete symmetries are called discrete gauge symmetries \cite{GGRoss92,Wilczek89,BanksDine92}.

\begin{figure}[!h]\hskip -0.3cm
\includegraphics[height=0.17\textwidth]{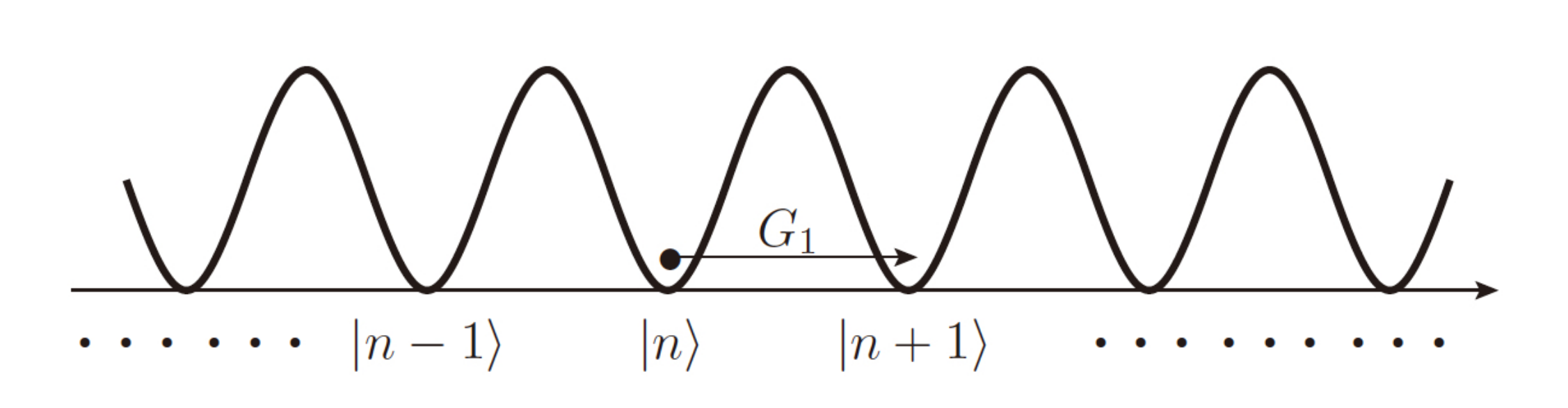}
\caption{Vacua $|n\rangle$ which change by one unit via the instanton gauge transformation $G_1$.}\label{fig:NonAVacua}
\end{figure}

Axion is an offspring from the discovery of nonlinear (instanton) solutions of non-Alelian gauge theories \cite{Belavin75}.  Instanton configurations are distinguished by the topological number called Pontryagin index
\dis{
q=\frac{1}{32\pi^2}\int d^4 x\,F^{a}_{\mu\nu}\tilde{F}^{a}_{\mu\nu}\label{eq:PonIndex}
}
where $\tilde{F}^{a}_{\mu\nu}=\frac12 \varepsilon_{\mu\nu\rho\sigma}F^{a\,\rho\sigma}$ is the dual of ${F}^{a}_{\mu\nu}$.
The simplest instanton (anti-instanton) solution \cite{Belavin75} gives $q=+1(-1)$.
Thus, gauge fields are distinguished by the Pontryagin index $^{(q)}\hskip -0.08cm A_\mu^a,$ with $q=0,\pm 1, \pm 2, \cdots$, and  \cite{Callan76}
\dis{
\frac{1}{32\pi^2}\int d^4 x\,^{(q)}\hskip -0.08cm F^{a}_{\mu\nu}\,  ^{(p)}\hskip -0.08cm \tilde{F}^{a}_{\mu\nu}=q \delta_{qp}.\label{eq:orthogonality}
}
Thus, different gauge field configurations or VEVs give different topological numbers. The instanton of Ref. \cite{Belavin75}  changes the Pontryagin index  by one unit,
\dis{
\int d^4 x\left( ^{(1)}\hskip -0.08cm F^{a}_{\mu\nu}+ ^{(n)}\hskip -0.08cm F^{a}_{\mu\nu}\right)&\left( ^{(1)}\hskip -0.08cm \tilde{F}^{a}_{\mu\nu}+ ^{(n)}\hskip -0.08cm \tilde{F}^{a}_{\mu\nu}\right)= \int d^4 x\left( ^{(1)}\hskip -0.08cm F^{a}_{\mu\nu}\, ^{(1)}\hskip -0.08cm \tilde{F}^{a}_{\mu\nu}+ ^{(1)}\hskip -0.08cm F^{a}_{\mu\nu}\, ^{(n)}\hskip -0.08cm \tilde{F}^{a}_{\mu\nu}+ ^{(n)}\hskip -0.08cm F^{a}_{\mu\nu}\, ^{(1)}\hskip -0.08cm \tilde{F}^{a}_{\mu\nu}+ ^{(n)}\hskip -0.08cm F^{a}_{\mu\nu} \, ^{(n)}\hskip -0.08cm \tilde{F}^{a}_{\mu\nu}\right)\\
&=32\pi^2(1+0+0+n)=32\pi^2(n+1).\label{eq:orthexample}
}
In Fig. \ref{fig:NonAVacua}, we show the vacua and the instanton changing the Pontryagin index by one unit. 

These $|n\rangle$ vacua can combine to define the so-called $\theta$ vacuum \cite{Callan76,Jackiw76},
\dis{
|\theta\rangle\propto~\sum_{n=-\infty}^{n=+\infty}~ |n\rangle e^{in\theta}.\label{eq:thetaVac}
}
The $|\theta\rangle$ vacuum is invariant under gauge transformations.

\subsubsection{Discrete gauge symmetries}

Continuous global symmetries can have discrete symmetries as subgroups and discrete sets of vacua. Cosmological effects of discrete vacua were first noticed by Okun \etal \cite{Okun74}. The boundaries of different regions of discrete vacua form domain wall during the cosmic evolution  \cite{Kibble77} and can lead to problems if decay of these domain walls produces too much radiation.  For the study of cosmic evolution, one needs an explicit model. For the QCD axion, Sikivie pointed out that the discrete axionic vacuum must be identity, \ie $\NDW=1$ \cite{Sikivie82DW}.

If that continuous symmetry is promoted to local symmetry,  then the different vacua in case of global symmetric case are connected by gauge transformations and the domain wall problem disappears. Looking at the set of discrete vacua  at low energy, can we confirm that there is no domain wall problem? References \cite{GGRoss92,Wilczek89} named this kind discrete symmetry {\it discrete gauge symmetry}.
 
The promoted local symmetry is designed to be broken at a GUT scale in the field theory GUT models or at   the compactification scale in models with extra dimensions. Thus, the magnitude violating the discrete gauge symmetry is suppressed by the  inverse powers of $M$(the GUT mass or the compactification scale).
In model building, these breaking terms are designed to be located in the allowed regions of the domain wall cosmology.

\subsubsection{$\NDW=1$ schemes}\label{subsubsec:SolDW1}

If the \UPQ~belongs to the center of a gauge group, then these QCD vacua are identified by a gauge transformation. The $N$ elements of $\Z_N$ are $\{1, e^{2\pi i/N}, e^{4\pi i/N}, \cdots,e^{2(N-1)\pi i/N} \}$. This can be identified with the elements of SU($N$) center,
\dis{
\frac{1}{\sqrt{2N}} \begin{pmatrix} 1&0&0&\cdots&0 \\ 0& e^{2\pi i/N}&0&\cdots&0 \\
 \cdots& \cdots& \cdots& \cdots& 0\\ 0&0&0&\cdots&e^{2(N-1)\pi i/N} 
 \end{pmatrix}  . }
 Since the SU($N$) elements are connected by gauge transformations, the elements of  $\Z_N$ are identified.
This kind of solutions is called the Lazarides-Shafi mechanism 
\cite{LShafi82}, and was applied in some SU(9) family unification GUT models \cite{Dimopoulos82,KKang83,KOuvry83}. To aquaint with the extended-GUT and tensor notations for the representations, we illustrate a solution given in Ref. \cite{Dimopoulos82}. The SU(9) indices are the lower case Roman characters, $a=\{1,2,3,\cdots, 9\}$. The multiple superscript (or subscript) indices below imply the antisymmetric combinations.
The Higgs representations are an adjoint $H^a_b(={\bf 80}), H^{abc}(={\bf 84})$,  $ H^{ab}(={\bf 36})$ and several $ H^{a}(={\bf 9})$. Because we are interested in breaking the PQ symmetry at the intermediate scale, we will omit the discussion on the fundamental representation needed for the Higgs fields. The L-hand chiral representations are the anomaly free combination $\psi^{abcd}\oplus \psi^{abc}\oplus 2\cdot \psi_{ab}\oplus 4\cdot \psi_a$. The anomalies of the antisymmetric representations of SU($N$) gauge groups are
\dis{
\rm SU(5)\hskip 3.4cm   ~\psi^a =+1,~&-1=\psi_a\\
\rm SU(6)\hskip 3.4cm    ~\psi^a = +1,~   &0,-1=\psi_a\\
\rm SU(7)\hskip 2.6cm    ~\psi^a = +1,+3,~2&,-2,-3,-1=\psi_a\\
\rm SU(8)\hskip 2.45cm    ~\psi^a = +1,+4,+5&,0,-5,-4,-1=\psi_a\\
\rm SU(9)\hskip 1.8cm    ~     \psi^a=+1,+5,+9,+5&,-5,-9,-5,-1=\psi_a\\
    \uparrow~~\uparrow~~\uparrow&~~\uparrow~~\uparrow~~~\uparrow\\
    \psi^{ab}~[3]~[4] & ~~\psi_{ab}\,\psi_{abc}\,\psi_{abcd}
    \label{eq:AnomSU}
} 
where the SU($N+1$) anomalies $A([n])$ are obtained by the triangular sum of  SU($N$) anomalies, and $[3]$ and $[4]$ in the last line of Eq. (\ref{eq:AnomSU}) mean antisymmetric combination of 3 and 4 indices, respectively. The numbers of quarks in the above representations {\bf 126}, {\bf 84},  {\bf 36}, and {\bf 9}  are 35, 21, 7, and 1, respectively. Thus, $A([4])+A([3])+2A([8])+4A([9])=+5+9-2\cdot 5-4\cdot 1=0$. For the Peccei-Quinn charges,   we assign $\QPQ(H^{ab})=+2$ and  $\QPQ(H^{abc})=-6$ from the Higgs coupling $\varepsilon_{abcdefghi}H^{abc}H^{de}H^{fg}H^{hi}$.
Possible Yukawa couplings allowed by the SU(9) gauge symmetry are
\dis{
\psi^{abcd}\psi_{ab}H_{cd},  \psi^{abc}\psi_c H_{ab},\psi^{abcd}\psi_d H_{abc},\varepsilon_{abcdefghi}\psi^{abcd}\psi^{efgh}   H^{i},  \psi_a\psi'_b H^{ab}\label{eq:allYuk}.
}
Note that $\varepsilon_{abcdefghi}\psi^{abc}\psi^{def}H^{ghi}$ is absent from the SU(9) symmetry.
If we allow all the Yukawa couplings given in Eq. (\ref{eq:allYuk}), it is not possible to introduce the PQ symmetry. To introduce the PQ symmetry, usually one exclude some terms allowed by the gauge symmetry, for which  usually discrete symmetries are introduced. To have a successful Lazarides-Shafi mechanism, we introduce only the first term and exclude the rest in  Eq. (\ref{eq:allYuk}). For example, we can introduce $\Z_3\times \Z_4$ symmetry with the assignments $\Z_3(H^{abc})=0,\Z_3(H^{ab})=1,\Z_3(H^{a})=2,\Z_4(\psi^{abcd})=0,\Z_4(\psi^{abc})=2,\Z_3(\psi_{ab})=1,$ and $\Z_3(\psi_{a})=-1$.
 Then, we have
\dis{
&\QPQ(H^{abc})=-6, \QPQ(H_{abc})=+6, \QPQ(H^{ab})=+2,\QPQ(H_{ab})=-2,\\
& Q(\psi^{abcd})=+3,Q(\psi^{abc})=-1, Q(\psi_{ab})=-1, Q(\psi_d)=-4,
}
Thus, the total PQ charge of the chiral fields is 
\dis{ 
1\times (+3)\times 35+ 1\times (-1)\times 21 + 2\times (-1)\times 7+ 4\times (-4)\times 1= 54 
}
which is a multiple of 9. Thus, by repeated SU(9) gauge transformations the axion $\Z_n$ vacua are connected.

Mathematically, the  Lazarides-Shafi mechanism requires that the domains of the PQ vacua $\Z_n$ rest in the center of the gauge group. The centers of continuous groups are \cite{Bacry77}
\dis{
 SU(N): \hskip 2.07cm  &\Z_N,~N\ge 2\\
 SO(2l+1):\hskip 1.5cm  &\Z_2,~l\ge 2\\
 SO(4k): \hskip 2cm  &\Z_2\times \Z_2,~k\ge 2\\
SO(4k+2):\hskip 1.4cm   & \Z_4 \\
Sp(2n):\hskip 2.1cm  &\Z_2,~n\ge 3\\
{\rm E}_6:\hskip 2.7cm  &\Z_3\\
{\rm E}_7:\hskip 2.7cm  &\Z_2\\
{\rm E}_8, {\rm F}_4, {\rm G}_2 :\hskip 1.55cm  &{\rm trivial}.\\
}
 
\begin{figure}[!t]\hskip -0.3cm
\includegraphics[height=0.55\textwidth]{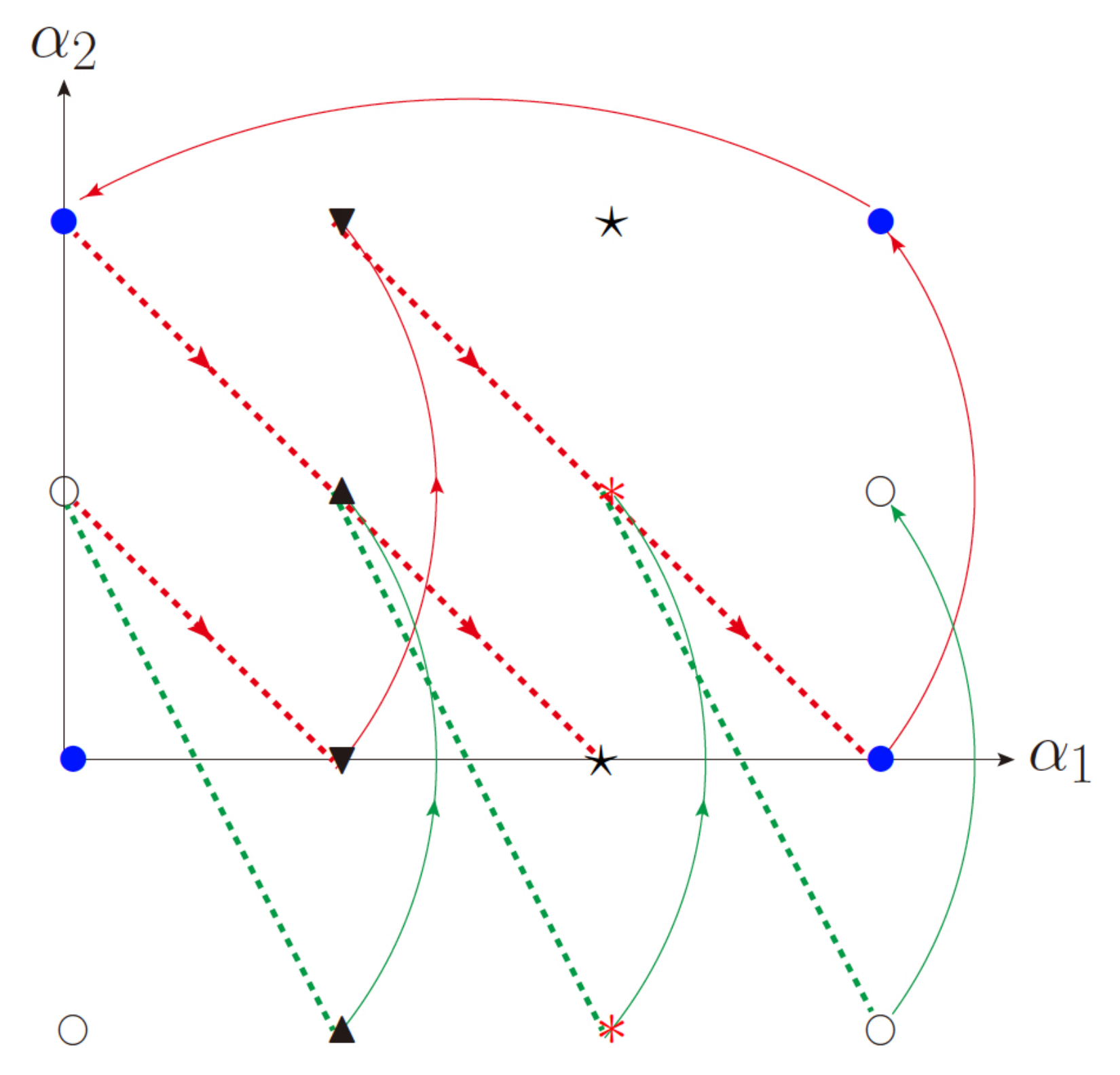} (a)
\caption{ The Choi-Kim mechanism with two Goldstone boson directions for $N_1=3$ and $N_2=2$ \cite{CKim85,KimNam21}. }
\label{fig:CKmechanism}
\end{figure}

Another interesting method is using two or more axions $a_i$ such that the discrete symmetries of the axionic vacua are connected, which can be called the Choi-Kim mechanism. The Choi-Kim mechanism is useful in multi-axionic models such as in string compactification \cite{CKim85}. In Fig. \ref{fig:CKmechanism}, we show the Choi-Kim mechanism with two Goldstone boson directions $\alpha_1=a_1/f_1$ and $\alpha_2=a_2/f_1$ \cite{CKim85}. Two  axionic discrete groups are $\Z_3$ from  $a_1$ and   $\Z_2$ from $a_2$. There are six independent vacua,
\dis{
  {\color{blue} \bullet}, \blacktriangledown, \star,  \circ , \blacktriangle, {\color{red}*} 
}
We can identify all six vacua by the red dashed lines which is the Goldstone boson direction. Notice, however, if we identify along the green dashed lines then only $\circ,\blacktriangle$ and ${\color{red}*} $ are identified. Shifting the green dashed lines one units upward, there is another identification of  $\bullet,\blacktriangledown$ and $\star$. Thus, there remains $\Z_2$ along the identification of green lines. This example shows that all the Goldstone boson directions are not necessarily identifying all the vacua. In the red dashed arrow case, both $\alpha_1$ and $\alpha_2$ are increased by one units while $N_1$ and $N_2$ are relatively prime. On the other hand, for the green dashed arrow case,
$\alpha_1$ is  increased by one unit while $\alpha_2$ is increased by two units. The two unit shift cannot distinguish $\Z_2$ and there remains the $\Z_2$ symmetry. Therefore, it is important to find relatively prime numbers for the shifts considering $\Z_N$'s.
Since there is the red dashed line identification, we conclude that all six vacua are identified. This depends on 2 and 3 of $\Z_2$ and $\Z_3$ being of relative prime.

\begin{figure}[!t]\hskip -0.3cm
\includegraphics[height=0.2\textwidth]{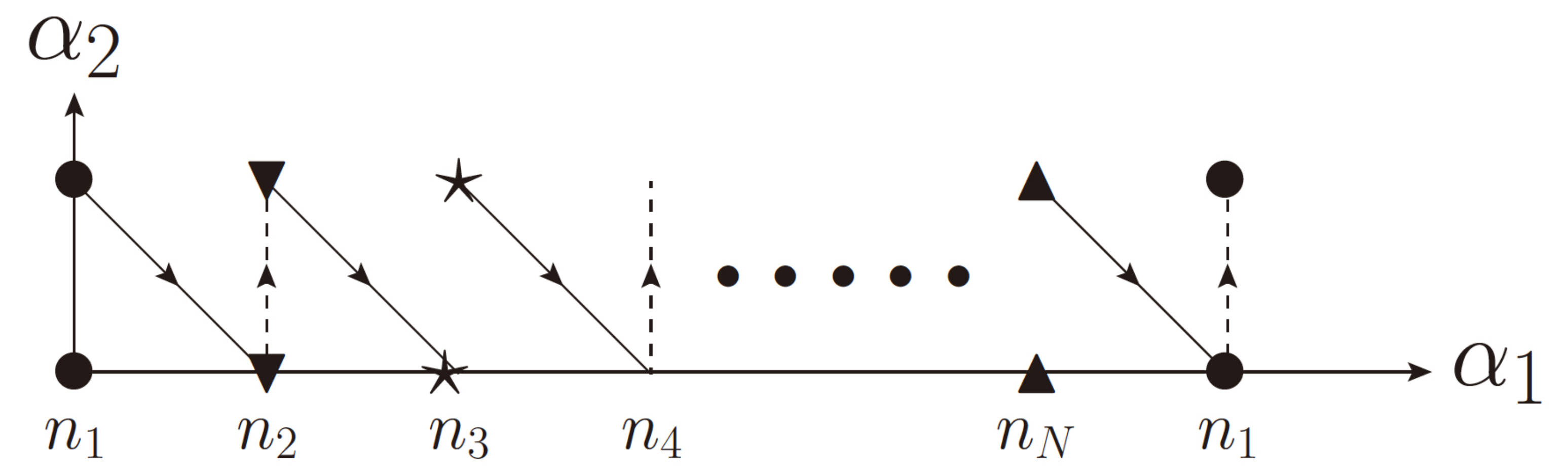}
\includegraphics[height=0.2\textwidth]{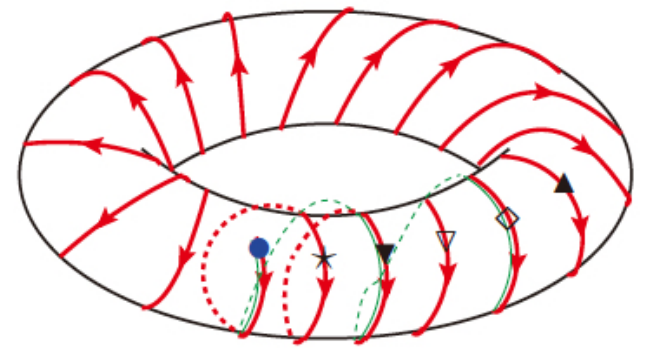}
\caption{ {\it (Upper fig.):} $\Z_N$ in $\alpha_1$ and $N_2=1$ in $\alpha_2$ directions.  {\it (Lower fig.):} The red curve is the example for $N=6$  \cite{KimNam21}. }
\label{fig:CKmech16}
\end{figure}

The above Choi-Kim mechanism has a far-reaching consequence if two discrete groups are $\Z_N$ and $\Z_1$  (\ie the same vacuum repreats in one direction). Since 1 and $N$ can be considered as  being relatively prime, all vacua are identified. This case is shown on the torus of Fig. \ref{fig:CKmech16}  for $N_1=6\, (\bullet, \star,\blacktriangledown,\triangledown,\diamond,\blacktriangle)$ and $N_2=1$.

\subsection{Quintessential axion}\label{subsec:ULA}

The word quintessence was used for a scalar field responsible for DE in the universe \cite{Steinhardt98}. An axion working for DE is a quintessential axion \cite{KN03}. It is a kind of ultra light axion (ULA) designed for DE of the universe. The current bounds on ULA interactions are presented in Ref. \cite{ChoiK21}.

Quintessential axion requires \cite{Carroll98}:
\begin{itemize}
\item The decay constant is near the Planck scale $\Mp$, and
\item Mass is about $10^{-32\,}\eV$.
\end{itemize}
Since the Planck mass is the mass defining scale, the first condition looks easy to be implemented. But, the BSM singlet $\sigma$ may be required to interact with light fields, in which case a judicious care is needed to allow the VEV of $\sigma$ still remaining near the Planck scale. Implementing the second condition needs some symmetries such that the breaking scale discussed in Subsecs. \ref{subsec:Vs}--\ref{subsec:Anomaly} are sufficiently small.

\begin{figure}[!t]\hskip -0.3cm
\includegraphics[height=0.55\textwidth]{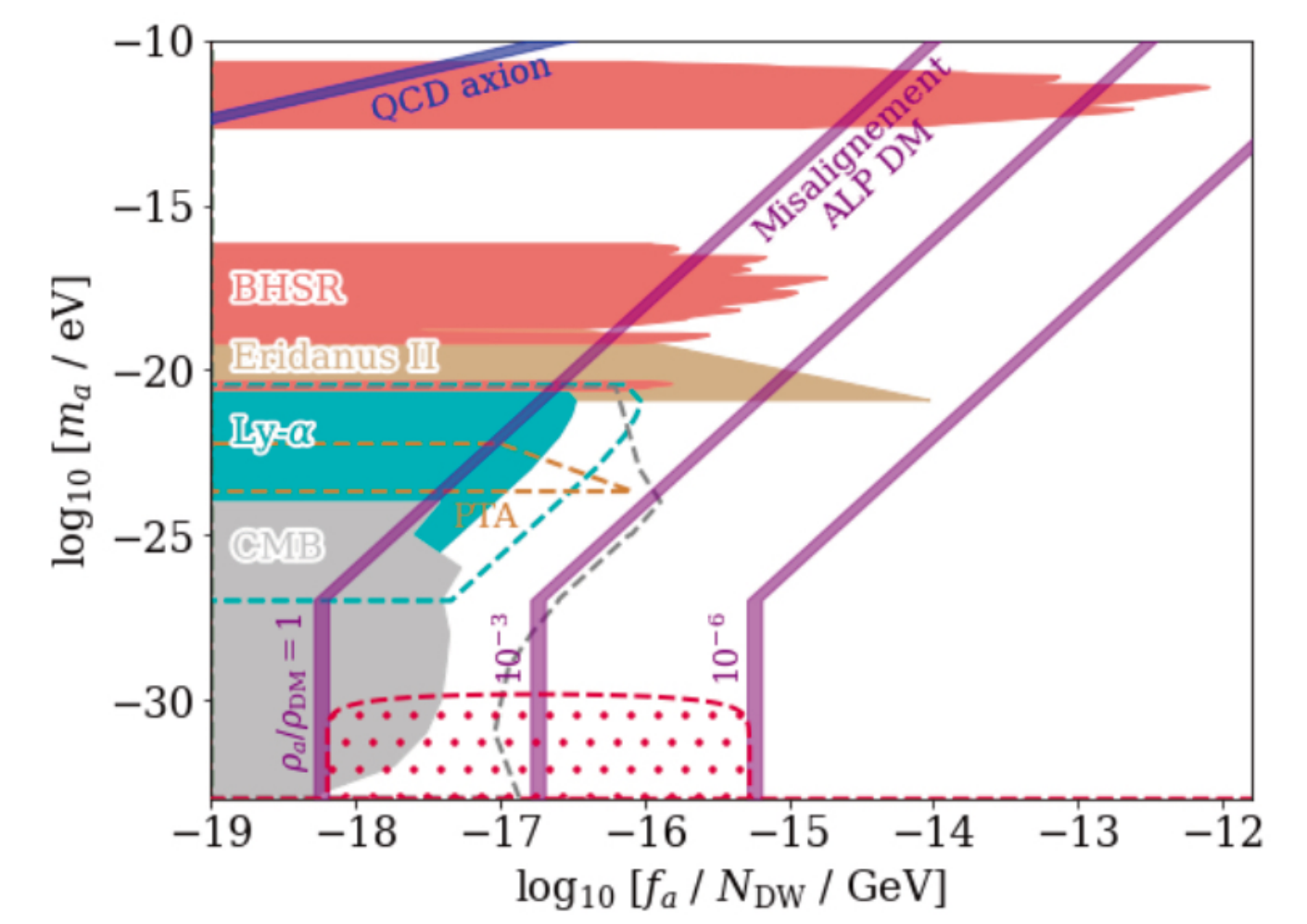}
\caption{ A summary of the
axion scale $f_a/\NDW$ versus axion mass from gravitational probes \cite{ChoiK21}. The shaded regions are excluded by the existing constraints, while the dashed lines show
the sensitivities of future experiments.  $f_a/\NDW$ is identified as the field VEV $\langle a\rangle$ for ALP DM or DE. }\label{fig:ULA}
\end{figure}

But it is a trend to call a scalar as an ALP also if its mass is sufficiently small even though the breaking is not by gauge anomalies. A summary of ULA limits is shown in Fig. \ref{fig:ULA}, where the region for the QA interest is shown with red dots \cite{KNP05}.

\subsubsection{QA through breaking by SU(2)$_W$ anomaly}

In the minimal supersymmetric standard model (MSSM),   the weak SU(2) gauge coupling constant is sufficiently small and the  SU(2)$_W$ anomaly can provide the needed small breaking of \UDE. For $  \alpha_{\rm em}(M_Z)=1/(127.940\pm 0.014)$, or for 
\dis{\frac{1}{\alpha_{2}(M_Z)}= 29.600\pm 0.010,
}
running gives the value at the GUT scale,  $1/\alpha_{GUT}$, as shown in Table \ref{tabLCMS15}  \cite{CMS1,CMS2}.  
In the MSSM, there already exists an excellent estimate,  based on the two-loop MSSM running of $\alpha_2$ \cite{Bourilkov15}.
\begin{table}[h!]
\begin{center}
\begin{tabular}{@{}|c| c c c|c|@{}} \hline
Threshold correction [\%] &$M_{SUSY}[\gev],$&  $Mg[\gev],$ &    $1/\alpha_{GUT}$& $\chi^2$\\[0.2em] \colrule &&&&  \\[-1.4em] 
 ~$+1$~  & $10^{3.96\pm 0.10}$  & $10^{15.85\pm 0.03}$ &   $26.74\pm 0.17$ &8.2\%   \\[0.3em]  \\[-1.4em] 
~$\pm 0$~  & $10^{3.45\pm 0.09}$  & $10^{16.02\pm 0.03}$ &   $25.83\pm 0.16$ &8.2\%   \\[0.3em]  \\[-1.4em] 
 ~$-1$~  & $10^{3.02\pm 0.08}$  & $10^{16.16\pm 0.03}$ &   $25.07\pm 0.15$   &9.5\%     \\[0.3em]  \\[-1.4em] 
 ~$-2$~  & $10^{2.78\pm 0.07}$  & $10^{16.25\pm 0.02}$ &   $24.63\pm 0.13$  &25.1\%      \\[0.3em]  \\[-1.4em] 
 ~$-3$~  & $10^{2.60\pm 0.06}$  & $10^{16.31\pm 0.02}$ &   $24.28\pm 0.10$   &68.1\%     \\[0.3em]  \\[-1.4em] 
 ~$-4$~  & $10^{2.42\pm 0.05}$  & $10^{16.38\pm 0.02}$ &   $23.95\pm 0.09$ &138.3\%   \\[0.3em]  \\[-1.4em] 
 ~$-5$~  & $10^{2.26\pm 0.05}$  & $10^{16.44\pm 0.02}$ &   $23.66\pm 0.09$   &235.7\%     \\[-0.1em]
 \hline
\end{tabular}
\end{center}
\caption{Combined fit results - CMS data \cite{CMS1,CMS2} and GUT unification.  The favored SUSY scale is 2820 GeV. Here, the threshold correction $\varepsilon_{GUT}$ at $\Mg$ is defined as $\alpha_3(\Mg)=\alpha_{GUT}(1+\varepsilon_{GUT})$.}\label{tabLCMS15}
\end{table}
For SU($N$) groups, we use the following $\beta$-function with gauge bosons and fermions \cite{Jones74, Caldwell74},
\dis{
\beta=-(\frac{\alpha_s}{4\pi})^2\left(\frac{11}{3}C_2(G)- \frac23\sum_R T(R)-\frac{\alpha_s}{4\pi}\Big(\frac{10}{3}\sum_R   C_2(G)T(R) +2\sum_R C_{2\ell}({\rm SU}(N))-\frac{34}{3}(C_2(G))^2\Big)\right)
}
where
\dis{
C_2({\rm SU}(N))=N,~C_{2\ell}({\rm SU}(N))=\frac{N^2-1}{2N}\ell(R),~T(R)=\ell(R), 
}
where $\ell({\bf N})=\frac12$. 
 Table \ref{tabLCMS15} was obtained with the following input parameters,
\dis{
 {\rm At }~ M_Z= 91.19\,\gev:~\left\{~\begin{array}{l}\sin^2\theta_W\Big|_{\overline{\rm MS}} = 0.23126 \pm 0.00005,\\[0.3em]
 \alpha_s=  0.1185 \pm 0.0006,\end{array} .\right.
}

 Without considering the threshold corrections at the GUT scale, we obtain   \cite{Bourilkov15},
\begin{eqnarray}
\textrm{MSSM}:&&\left[ \begin{array}{l}
e^{-2\pi/\alpha_2}\Big|_{\Mg}=1.69\times 10^{-81},\\[2em]
\left\{~\begin{array}{l}M_{\rm SUSY}=2820+670-540~\gev,\\[0.3em]
\Mg=(1.065\pm 0.06) \times 10^{16}\gev.\end{array} \right.
\end{array}\right.\\[1em]
\textrm{SM}:&&\left[ \begin{array}{l}
e^{-2\pi/\alpha_2}\Big|_{\Mg}=1.69\times 10^{-131},\\[1em]
\Mg=(1.096\pm 0.06) \times 10^{15}\gev.\end{array}\right. \label{eq:SMefactor}
\label{eq:MSSM}
\end{eqnarray}
Note that the SU(2) coupling does not run with one-loop corrections in the MSSM. On the other hand, in the non-supersymmetric SM, the SU(2) coupling runs and we obtain a much smaller factor if not considering the threshold corrections. If the SU(2) gauge anomaly is responsible for DE, we require the following order for $\Lambda$ in the MSSM,
\dis{
\textrm{MSSM}:~1.69\times 10^{-81}\Lambda^4=(0.003\,\eV)^4\to \Lambda\sim 1.48\times 10^{8}\,\gev,\\
\textrm{SM}:~1.065\times 10^{-131}\Lambda^4=(0.003\,\eV)^4\to \Lambda\sim  5.25\times 10^{20}\,\gev.\label{eq:SU2scale}
}
Thus, in the SM we cannot use the SU(2) gauge anomaly  for the source of DE. But, in the MSSM  the SU(2) gauge anomaly can work for the explicit breaking term for the QA. If it oscillated before the current epoch, then it cannot work for DE but may work for a ULA.

\subsubsection{By potential terms}

The general form for explicit breaking terms in the potential, $\Delta V$, is given  in Eq. (\ref{eq:ExBreakingV}), 
\dis{
\Delta V= &\left(\frac{1}{(M)^{2i+j+2l-4}}(\sigma^*\sigma)^i\sigma^j +\frac{1}{(M)^{2i'+j'+2l-4}}(\sigma^*\sigma)^{i'}\sigma^{*j'} \right) (H_u H_d)^l+{\rm h.c.} 
\label{eq:BRbyV}
}
where $i,j,i',j',l =\rm integers$. The leading term with $i=i'=l=0$ and $j=j'=1$ was used for the QCD axion in Ref. \cite{Holman92,Barr92}, where terms $\sigma^n$ are required to be forbidden up to $n=8\sim 9$.

\subsubsection{QA through breaking by Yukawa couplings}

\begin{figure}[!t]\hskip -0.3cm
\includegraphics[height=0.25\textwidth]{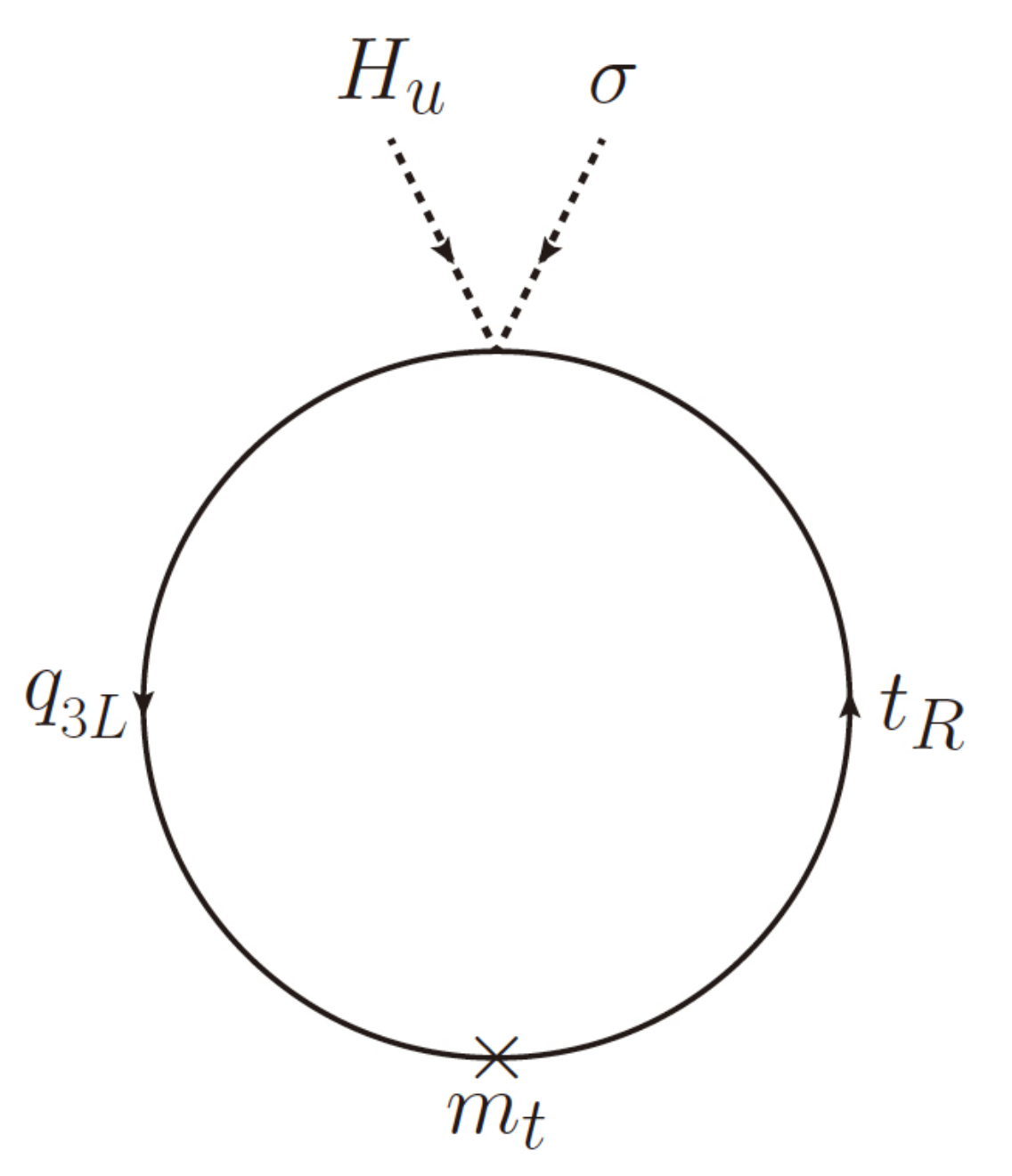}
\caption{\label{fig:YukawaLoop} Potential generated by Yukawa terms breaking \UDE.}
\end{figure}

In the SM with one Higgs doublet $H_d$, there is no PQ symmetry. Yukawa couplings for the up-type quarks are given by $\bar{q}_Lu_R \tilde{H}_d $ in terms of $\tilde{H}_d=F{H}_d^*$ with the flipping matrix 
\dis{
F=\begin{pmatrix} 0&-1\\+1&0\end{pmatrix}.
}
The BSM field $\sigma$ multiplied to the SM Yukawa couplings explicitly break the \UDE~symmetry, defined for the scales much below $M_d$ and $M_u$,
\begin{eqnarray}
\Delta {\cal L}_Y=&\frac{1}{M_d}\bar{q}_Ld_R H_d\,  \sigma,\cdots\label{eq:BreakYd}\\[0.3em]
&\frac{1}{M_u}\bar{q}_Lu_R \tilde{H}_d \, \sigma,\cdots\label{eq:BreakYu}
\end{eqnarray}
For a loop diagram Fig. \ref{fig:YukawaLoop} generated by Eq. (\ref{eq:BreakYu}), we obtain a linear term for $\sigma$ as
\dis{
\Delta V_Y\approx \frac{\Lambda^2}{16\pi^2}\frac{m_{\rm top}}{M_u}\langle \tilde{H}_d\rangle \sigma
+{\rm h.c.}\label{eq:LoopV}
}
where $\Lambda^2$ is the cut-off scale in the loop integral. It is the scale where new physics takes over the SM.
Generally the generated term should be less than the DE magnitude of order $(0.003\eV)^4$.
For an intermediate value for $\sigma$ of order $\approx 10^{10}\gev$, $M_u$ is  required to be greater than $3\times 10^{64}\cos\beta\gev$ for $\Lambda=$TeV. So, the linear terms given in Eqs. (\ref{eq:BreakYd},\ref{eq:BreakYu}) are forbidden.   In supersymmetric models also, this conclusion is valid. So, the intermediate scale for $\sigma$ breaking \UDE~symmetry via Yukawa couplings is forbidden.  However, if the linear terms given in   Eqs. (\ref{eq:BreakYd},\ref{eq:BreakYu}) respect the \UDE~symmetry, then they are allowed.

Therefore, if Yukawa couplings break the \UDE~symmetry, the breaking scale via $\sigma$ should be at a very low energy scale.  However, it is against the requirement for a quintessential axion suggested in the beginning of this Subsec. From an esthetic viewpoint, therefore, breaking by potential terms or Yukawa couplings are not very attractive in simple models with one BSM field $\sigma$.
 
\subsection{QCD axion plus QA/ULA}

This leads us to multi-$\sigma$ fields. A more realistic two U(1) symmetries are \UPQ~and \UDE. It is more likely that up-type quarks and down-type quarks carry different charges under some U(1). In this case, if there is no family dependence, we can choose that U(1) as \UPQ.  Thus, with two U(1) symmetries, we can obtain  in most cases \UPQ~and \UDE.

Since we consider \UPQ\, to admit the axion solution of the strong CP problem \cite{KimRMP10}, its breaking is by the QCD anomaly with no breaking or sufficiently suppressed  breaking by the potential. For breaking \UDE, let us consider first the SU(2) anomaly and next a sufficiently suppressed potential terms. 

\subsubsection{SU(2) anomaly}

 In Eq. (\ref{eq:SU2scale}), $\Lambda$ for SU(2) anomaly toward QA was calculated as $1.48\times 10^8\,\gev$. Note that, among the terms breaking global symmetries  the QCD anomaly is the largest one and QA/ULA must be chosen after removing the QCD anomaly part.  To house the BSM pseudo-Goldstone bosons with \UPQ~and \UDE, we introduce two $\sigma$ fields,   $\sigma$ and $\sigma_{\rm quint}\equiv \sigma_q$ whose VEVs are
\dis{
\langle \sigma\rangle=\frac{f_a}{\sqrt2}\, e^{ia/f_a},~
\langle \sigma_{\rm quint}\rangle=\frac{f_q}{\sqrt2}\, e^{ia_q/f_q}. 
}
  $\sigma$ and $\sigma_q$ carry  the following  U(1)$_{\rm PQ}$ and \UDE\, charges,
\dis{
\begin{array}{ccc}
&~~\sigma~~ &\sigma_q\\
\QPQ: & 1 & \Gamma_2\\
\QDE : & \Gamma_1 & 1
\end{array}
} 
Then,  the SU(3) and SU(2) instantons generate the following potential,
\dis{
V= m\Lambda^3_{\rm QCD}\left(\cos(\frac{a}{f_a} +\Gamma_2\frac{a_q}{f_q} )+{\rm h.c.}\right)+f_q^4e^{-2\pi/\alpha_2}\left(\cos(\Gamma_1\frac{a}{f_a}+\frac{a_q}{f_q}) +{\rm h.c.}\right) \label{eq:TwoCos}
}
where $m$ is a typical chiral mass for light quarks, most probably very close to $m_u$.
The mass matrix from Eq. (\ref{eq:TwoCos}) is
\dis{
\begin{pmatrix}
\frac{m\Lqcd^3+\Gamma_1^2 f_q^4 e^{-2\pi/\alpha_2}}{f_a^2},& \frac{\Gamma_1 f_q^4 e^{-2\pi/\alpha_2}+\Gamma_2 m\Lqcd^3  }{f_af_q} \\[0.7em]
 \frac{\Gamma_1 f_q^4 e^{-2\pi/\alpha_2}+\Gamma_2 m\Lqcd^3  }{f_af_q} ,&
 \frac{f_q^4 e^{-2\pi/\alpha_2}+\Gamma_2^2 m\Lqcd^3}{f_q^2} 
\end{pmatrix}\label{eq:qMassM}
}
The eigenvalues of (\ref{eq:qMassM}) are  
\dis{
m^2_{a,a_q}=&\frac{1}{2 f_a^2 f_q^2} \Biggl(  f_q^4e^{\frac{-2\pi }{\alpha _2}}(f_a^2+f_q^2\Gamma_1^2)+
m\Lqcd ^3 (f_q^2+f_a^2\Gamma_2^2) \\
\\&\mp   \sqrt{-4(\Gamma_1\Gamma_2 -1)^2\,m \Lqcd^3   {f_a^2}{ f_q^6}e^{\frac{-2 \pi }{\alpha _2}} +\left(f_q^4e^{\frac{-2\pi }{\alpha _2}}(f_a^2+\Gamma_1^2f_q^2 ) +m\Lqcd^3(f_q^2+f_a^2\Gamma_2^2)\right)^2 }\,\Biggr)
}
If we neglect the $e^{-2\pi /\alpha _2}$ factor, two eigenvalues are
\dis{
m_a^2 &=\frac{m\Lqcd ^3}{ f_a^2 } \Biggl(1+\frac{\Gamma_2^2 f_a^2}{f_q^2}  \Biggr)+O(e^{\frac{-2\pi }{\alpha _2}}  ),\\
m_{a_q}^2&=  (\Gamma_1 \Gamma_2-1)^2f_q^2e^{\frac{-2\pi }{\alpha _2}}   \frac{1}{1+\Gamma_2^2 (f_a^2/f_q^2)} +O(e^{\frac{-4\pi }{\alpha _2}} ) \, .
}
Note that $e^{-2\pi /\alpha _2}$ is almost $0.169\times 10^{-40}$. With $f_q\simeq 10^8 \gev$, we obtain $m_{a_q}\simeq 2\times 10^{-13}\gev\approx 0.0002\eV$ and the vacuum energy density $f_q^4 e^{-2\pi /\alpha _2}\approx    (0.64\times 10^{-3}\eV)^4$. These $f_q\simeq 10^8 \gev$ and  $m_{a_q}\approx 2\times 10^{-4\,}\eV$, located in the allowed region in Fig. \ref{fig:ULA}, do not belong to the items listed in the beginning of  
Subsec. \ref{subsec:ULA}. Note, however, that the items listed there are anticipated from breaking by terms in the potential, $\Delta V$.

\subsubsection{By $\Delta V$}

To suppress terms to a high power, one generally use discrete symmetries. This suppression must be more effective than those considered in  Ref. \cite{BarrSeckel92,Kamionkowski92,Holman92} because we anticipate the VEV of $\sigma_q$ is near the Planck scale while the scale considered in   Ref.  \cite{BarrSeckel92,Kamionkowski92,Holman92}  is  the intermediate scale. One example was presented in Ref. \cite{KimPRD16} where suppression was based on the detail structure of the model. In general, this kind of detail is needed to guarantee a sufficient suppression. Since the VEV of $\sigma_q$ is near the Planck scale, suppression by  $\sigma_q$ alone is not very helpful. So, the discrete symmetry should allow always sufficiently large nonzero $l$ among  indices in consideration in Eq. (\ref{eq:BRbyV}) so that the electroweak scale VEVs appear with a high power.
Naively, we have 
\dis{
(2.43\times 10^{27}\eV)^4\cdot\left(\frac{246\gev}{2.43\times 10^{18}\gev}\right)^{2l} <(0.003\eV)^4 ,
}
gives $ l>3.7$, leading to   $l\ge 4$.
In Table \ref{tabtwoDs}, we show two examples of discrete symmetries where the charges of \UPQ\, and \UDE\, are  also shown. In SUSY models, more elaborate discrete charges can be given.

\begin{table}[h!]
\begin{center}
\begin{tabular}{@{}|c| c c cc|ccc|@{}} \hline
  &  ~$\sigma,$ &    $\sigma_q,$& $H_u,$& $H_d~$ & ~$q_L,$& $\bar{u}_L,$& $\bar{d}_L$~\\[0.2em] \colrule &&&&  \\[-1.4em] 
 ~$\Z_4$~  & ~$0$  & $0$ &   $0$ &1 &0 &0 &3~ \\[0.3em]  \\[-1.4em] 
~$\Z_8$~  & ~$0$  & $0$ &   $0$ &1 &0 &0 &7~\\[0.3em]
~$\QPQ$~  & ~1  & --1 &   $0$ &1  &0 &0 &--1~~\\[0.3em]  \\[-1.4em] 
~$\QDE$~  & ~--1  & 1 &   $0$ &1&0 &0 &--1~~\\[0.1em]
 \hline
\end{tabular}
\end{center}
\caption{Two discrete symmetry examples.}\label{tabtwoDs}
\end{table}

\subsubsection{Cosmological effects}

The QA cosmology  depends on the cosmic time scale of $\sigma$ domination and the lifetime of $\sigma$.  Toward a search of these time scales, an explicit model is required. Here, we note that even if some terms in the potential are sufficiently suppressed, effective derivative interactions may contain terms affecting the QA scenario. If string compactification is used, moduli from string compactification may have dominant effects but it is impossible to draw model-independent results. Since string moduli have axionphilic nature owing to the model-insensitive derivative
interactions arising from the  K$\ddot{\rm a}$hler  potential, the best we can draw is the axionphilic nature derived from these derivative interactions that is discussed in \cite{ParkWI21}, which is summarized in the subsequent paragraph. 

The decay of a modulus into stringy axions occurs without
suppression by the mass of final states. Interestingly, it turns out to hold in general not only for the scalar
partner of the stringy axion but also for any other moduli. The decay into (pseudo)-Nambu-Goldstone
bosons (NGBs) also avoids such mass suppression if the modulus is lighter than or similar in mass to the
scalar partner of the NGB. Such axionphilic nature makes string moduli a natural source of an observable
amount of dark radiation in string compactifications involving ultralight stringy axions, and possibly in
extensions of the Standard Model that include a cosmologically stable NGB such as the QCD axion. In the
latter case, the fermionic superpartner of the NGB can also contribute to the dark matter as QCD axino.

In supersymmetric extension, L-handed chiral fields are represented by complex fields. Therefore, 
let us consider two complex singlets, $\phi(x)$ and $\varphi(x)$, which contain NGBs for the QCD axion $a$ and the quintessential axion $\varepsilon$,
\dis{
\phi(x)&=\{\sigma, a\}=\frac{f}{\sqrt2}\left(1+\frac{\sigma(x)}{f} \right) e^{i\,a(x)/f},\\
\varphi(x)&=\{d,\varepsilon\}=\frac{1}{\sqrt2}\left(d(x)+i\varepsilon(x)\right).\label{eq:compsinglets2}
}
Then, we follow the supergravity discussion of Ref. \cite{ParkWI21} for picking up the derivative interactions.
The  K$\ddot{\rm a}$hler  potential modified by modulus $\varphi$ is included in
\dis{
K_0+Z\Phi\Phi^*
}
where $Z$ is the field dependent coupling and the superfield $\Phi$ is
\dis{
\Phi=\varphi +\sqrt2 \theta \psi + \theta\theta F.
}
We set the reduced Planck mass $\Mp$ at 1. The shift of modulus $\delta S=S-S_0$ derives the energy shift  from the equilibrium point. In the real  K$\ddot{\rm a}$hler  potential, the effect is
appearing as mixing terms
 \dis{
 (\delta S+\delta S^*)&\Phi\Phi^*\big|_{D\,\rm term}=\delta S \Phi\Phi^*\big|_{D\,\rm term}+ H.c.\\[0.3em]
\to fd\,&\partial^2\sigma+f\varepsilon\,\partial^2a+ H.c.\label{eq:MixingInt}
 }
and the scalar cubic interaction terms
 \dis{
 (\delta S+\delta S^*) \Phi\Phi^*\big|_{D\,\rm term} 
\to -\frac12 aa\, \partial^2d+a\sigma  \partial^2\varepsilon+  (d\sigma-\varepsilon a )\partial^2\sigma  + (da+\varepsilon \sigma)\partial^2a  .\label{eq:CubicInt}
 }
 The overall coupling constant is determined by
\dis{
\kappa=-\frac{1}{\sqrt2}\langle (\partial^2K_0)^{-1/2}\partial_S\ln Z\rangle
} 
which is generally of order unity.

 \section{R parities in SUSY models}\label{sec:Rparity}
 
Another needed discrete symmetry is the R-parity in SUSY models.  We will use only L-hand chiral fields for writing the superpotential $W$. Because of the integration with the anticommuting parameter $\vartheta$, $\int d\vartheta^2 W$, the superpotential $W$ carries two units of $R$ charge of the global symmetry \UoR. If \UoR\, is broken to $\Z_n$, then the resulting discrete symmetry is called $Z_{n\,R}$. Customarily, R-parity is the case for $n=2$. For chiral fields, we use the upper case letters, e.g. the Higgs superfield is $H=h +\tilde{h}\vartheta +h_F\vartheta^2$ . Without  R-parity, $W$
admits the dimenesion 3 superpotential $U^cD^cD^c$ which breaks the baryon number since both $U^c$ and $D^c$ carry baryon number $-1$. With R-parity, such a term can be forbidden  \cite{Hall83}, since $U^c D^cD^c$ carries  $Q_R=3\ne 2$ with $Q_R(U^c)=Q_R(D^c)=1$ where $Q_R$ is the U(1)$_R$ quantum number. If we assume $\ZR$, we can consider only two integers 1 and 0, since all integers are included in 0 modulo 2 and 1 modulo 2.

In the minimal supersymmetric standard model (MSSM), the gauge invariant superpotential contains
\dis{
W=&+Y^{ij}_e L_i\overline{E}_j H_d+Y^{ij}_d Q_i\overline{D}_j H_d+Y^{ij}_u Q_i\overline{U}_j H_u\\
&+\mu H_uH_d+\kappa_i  H_u L_i\\
&+\lambda_{ijk}^{(0)} L_iL_j \overline{E}_k+ \lambda_{ijk}^{(1)} L_iQ_j \overline{D}_k + \lambda_{ijk}^{(2)} \overline{U}_i\overline{D}_j \overline{D}_k\\
&+\kappa_{ij}^{(0)} H_uL_i H_uL_j +\kappa_{ijkl}^{(1)} Q_iQ_jQ_k L_l   
+\kappa_{ijkl}^{(2)} \overline{U}_i\overline{U}_j \overline{D}_k\overline{E}_l\\
&+\kappa_{ijk}^{(3)} Q_iQ_jQ_kH_d+\kappa_{ijk}^{(4)}Q_i\overline{U}_j \overline{E}_kH_d+\kappa_{ijk}^{(5)}L_i H_uH_uH_d. \label{Wmssm}
}
With $\ZR$, the SM Yukawa couplings are given in the first line of $Y$ couplings.
All the other terms of SM fields in Eq. (\ref{Wmssm}) appear due to the extension to SUSY. The $\lambda$ and $\kappa$ terms break the lepton and/or baryon numbers.
The  $\ZR$ R-parity assignments of the MSSM fields which allow the $Y$ couplings are
\dis{
\begin{array}{cccccccc}
&~~Q~~ &~~\overline{U}~~&~~\overline{D}~~&~~L~~&~~\overline{E}~~&~~H_u~~&~~H_d~~\\
{\rm R-parity}: & -1 & -1 & -1 & -1 & -1 & +1 & +1 
\end{array}\label{eq:RpMSSM}
} 
which allow the $\mu$ and $\kappa^{(0)}$ terms also. In Eq. (\ref{eq:RpMSSM}), matter superfields have the minus R-parity while the Higgs superfields have the plus R-parity, or equivalently one and zero units of $\ZR$, respectively.
Equation (\ref{eq:RpMSSM}) can be called matter parity where Higgses belong to matter also. In a sense, matter fields survive below the GUT scale, and it is helpful if non-zero R-parity helps to make them chiral fields at the GUT scale.
  
 The $\mu$ term in Eq. (\ref{eq:RpMSSM}) is the needed one at the TeV scale for supersymmetry breaking. The $\kappa_{ij}^{(0)}$ term is allowed, breaking the lepton number in two units, which is the so-called ``Weinberg operator'' for neutrino masses \cite{Weinberg79}. But, the $\kappa_{ijkl}^{(1)}$ and $\kappa_{ijkl}^{(2)}$ terms, breaking the baryon number, are also allowed. Even though they are dimension-4 superpotential, the coefficients must be of order $10^{-7}$. Therefore, it is better to forbid the  $\kappa_{ijkl}^{(1)}$ and $\kappa_{ijkl}^{(2)}$ terms, and Ref. \cite{Lee11} realized it by a $\ZZR$ symmetry. The R-symmetry is defined by the shift of $\vartheta$, and $\ZZR$ shift is $\vartheta\to e^{2\pi i/4}\vartheta=i\vartheta$. Since $\int d^2\vartheta \,W$ must be invariant, the $\ZZR$ symmetry requires
\dis{
\vartheta &\to i\vartheta\\
W &\to -W. \label{eq:Z4R}
}
Equation (\ref{eq:Z4R}) is satisfied if we assign the following U(1)$_R$ quantum numbers, 
\dis{
\begin{array}{cccccccc}
&~~Q~~ &~~\overline{U}~~&~~\overline{D}~~&~~L~~&~~\overline{E}~~&~~H_u~~&~~H_d~~\\
\ZZR: & i & i & i & i & i & +1 & +1 
\end{array}\label{eq:Z4MSSM}
} 
The  $\kappa_{ijkl}^{(1)}$ and $\kappa_{ijkl}^{(2)}$ terms remain the same, \ie $W\to W$, under $\Z_{4R}$ and hence are forbidden. Also, the $\mu$ term is forbidden. However, the required $\mu$ term of O(TeV) can be induced at the GUT or Planck scale as a dimension-4 superpotential \cite{KimNilles84}. The allowed  dimension-4 superpotential terms of Ref.  \cite{KimNilles84} admit a global symmetry \UPQ, with the PQ charges of $\QPQ(H_u)=+1,\QPQ(H_d)=+1,\QPQ(Q_L)=-1, \QPQ(\overline{U}_L)=\QPQ(\overline{D}_L)=0$. For the PQ charges of leptons, we can assign any. For example, we can take the DFSZ assignment, $\QPQ(\ell_L)=-1, \QPQ(\overline{E}_L)=0$.  But,  the resulting PQWW axion must be forbidden \cite{KimRMP10}. Therefore, let us introduce a BSM singlet $\sigma$, with $\Z_{4R}(\sigma)=i$ with $W\to -W$, which allows a dimension-4 superpotential term
\dis{
W_\mu=c_{\mu} \frac{\sigma\sigma}{\Mp} H_uH_d.\label{eq:IntMu}
}
The superpotential $W_\mu$ also admits the PQ symmetry with the PQ charge of the BSM singlet $\QPQ(\sigma)=-1$. At the intermediate scale of the $\sigma$ VEV, the QCD axion is created \cite{KimPRL79,Shifman80,DFSa,Zhit}, and the resulting $\mu$ term is at the TeV scale \cite{KimNilles84},   for $\langle \sigma\rangle=1.5\times 10^{9\sim 12}\gev$,
\dis{
W_\mu \sim c_\mu\,\frac{(1.5\times 10^{9\sim 12}\gev)^2}{2.43\times 10^{18\,}\gev }\approx c_\mu\cdot(\gev\sim 10^3\tev).
}
A TeV scale $\mu$ term is needed to obtain the TeV scale superpartner masses. Note that the VEV of $\sigma$ breaks one global symmetry, a combination of U(1)$_R$ and \UPQ. Below the intermediate scale,  U(1)$_R$ is broken explicitly, and hence we introduce another global U(1) symmetry for a quintessential axion.

Introducing \UDE, we may assign non-vanishing \UDE~charges to leptons while vanishing \UDE~charges are to quarks. However, this scheme does not work since $(H_uH_d)$ carries  $\Z_{4R}=+1$ charge and there is no reason that terms up to $(\rm BSM~singlets)\cdot(H_uH_d)^3$ are suppressed to guarantee a sufficiently light quintessential axion. Even though we need details of more elaborate models for a successful introduction of \UDE, more discrete symmetries seem to be needed for  a sufficiently light quintessential axion.

Successful two discrete symmetries are given in a flipped-SU(5) GUT in Ref. \cite{KimPLB21}. There, its essence is presented in a next-minimal supersymmetric standard model (NMSSM) with two BSM singlets $\sigma$ and $\sigma_q$. We present the model in the following Eq. (\ref{eq:twoZNMSSM}) that leads vanishing  anomalies. Equation (\ref{eq:twoZNMSSM}) shows an example with no global anomalies of \UPQ$^3$, \UDE$^3$, \UPQ$^2$\UDE, and \UPQ\UDE$^2$,
\dis{
\begin{array}{cccccccccc||cc|ccccc}
&~Q~ &~\overline{U}~&~\overline{D}~&~L~&~\overline{E}~&~Q_H&~\overline{Q}_H&~H_u&~H_d&~\sigma~&\sigma_q&2\cdot \sigma_{\rm in1},&3\cdot \sigma_{\rm in2},&\sigma_{\rm in3},&2\cdot \sigma_{\rm in4},&  \sigma_{\rm in5}\\
\Z_{4R}~\rm charge: &+1 &+1 & +1 & +1& + 1 &  0&  0  & +2 & +2 & 0 & 0& 0& 0 & 0 & 0& 0   \\[0.3em]
\Z_2~\rm charge: & 0 &  0 &  0 &  0 &  0 &+1 & -2 & +1 & +1 & +1 & +1 & 0 & 0 & 0 &0& 0
\\[0.3em]
Q_{\rm PQ}: & + \frac12 &  + \frac12 &  + \frac12 &  + \frac12 &  + \frac12 & - \frac12 & - \frac12 & - 1 & - 1 & +1 &  0&-1 &+\frac12  &+1 &+\frac12&0
\\[0.3em]
Q_{\rm DE}: & +\frac12 &  +\frac12&  +\frac12 &  +\frac12 &  +\frac12 & 0 & 0 & -1 & -1 &  0 & -1 & -1& +\frac12 &0&0&+1
\end{array}\\ \label{eq:twoZNMSSM}
} 
 where three families of $Q, \overline{U}, \overline{D}, L, \overline{E}, Q_H$ and $\overline{Q}_H$ are introduced. The RHS of double vertical bar lists possible SM singlet fields.  From the consideration of different cosmological domains in the PQ vacuum,  Ref. \cite{Wilczek89} requires no global anomalies.
From the quarks listed in Eq. (\ref{eq:twoZNMSSM}), we have the axionic DW number 3. But the center of SU(3)$_c$ color group is 3 and these three axionic vacua are identified,  and hence there is no domain wall problem as discussed in Subsubsec. \ref{subsubsec:SolDW1} by the so-called Lazarides-Shafi mechanism \cite{LShafi82}.

 \section{String compactification}\label{sec:String}

To realize \UDE, in this section we adopt a top-down approach based on a string compactification listing full spectra of chiral fields. Difference in the L- and R-hand chiral fields are encoded at the compactification (to 4D) scale of our heterotic string \cite{ghmr84L,ghmr86np}. Below, the chirality will be denoted as $\oplus$ for R-fields and $\ominus$ for L-fields. But, a judicious compactification can lead to chiral spectra in other string models also. For an explicit example below, we use the orbifold model presented in \cite{KimPRD21}, 
\dis{
V_0=\left(\frac{1}{12},\frac{1}{12},\frac{1}{12},\frac{1}{12},\frac{1}{12},\frac{1}{12},\frac{1}{12},\frac{1}{12},\frac{1}{12};\frac{3}{12},\frac{6}{12}; \frac{6}{12}, \frac{6}{12}, \frac{6}{12}, \frac{6}{12}, \frac{6}{12} \right)\\
a_3=a_4=\left(0^9;\frac{-2}{3},0;~ \frac{2}{3},~ \frac{2}{3},~ \frac{2}{3},~ \frac{2}{3},~ \frac{2}{3}\right)\label{eq:OrbModel}
}
from which three twisted sectors are $V_0, V^+(=V+a_3),$ and $  V^-(=V-A_3)$. The model, Eq. (\ref{eq:OrbModel}), compactyifies SO(32) down to SU(5)$\times$SU(9)$'\times$U(1)$^4$. The sum of ranks of SU(5) and SU(9)$'$ is 12. Since the rank of SO(32) is 16, there remains (U(1)$)^4$ below the compactification scale. The chiral families are, 
three SU(5) families in the visible sector after removing vector-like pairs,
 \begin{eqnarray}
 3\cdot {\tenb}_L(T_7^+)+3\cdot {\five}_L(T_7^-),\label{eq:FFive}
\end{eqnarray}
shown in Table \ref{tab:T7minus}, 
and one SU(9)$'$ hidden sector family after removing vector-like pairs,
\begin{eqnarray}
 \nine'_L(U)+{\tsix}'_L(T_9^0)+[{\tsix}'_L(T_9^0)+{\tsix}'_R(T_9^0)] +7\cdot \nine'_R(T_4^0) 
  +7\cdot \nine'_L(T_4^-)\\
    +3\cdot \nine'_R(T_1^-) 
  +  3\cdot \nine'_R(T_2^0)+ 3\cdot \nine'_R(T_7^0)+ 3\cdot \nine'_L(T_7^+). \label{eq:FNine}
 \end{eqnarray}
 In Eq. (\ref{eq:FFive}), SU(5) families appear for three times ($\tenb+\five$) for which the SU(5) anomaly is $3(-1+1)=0$ from  Eq. (\ref{eq:AnomSU}).
In Eq. (\ref{eq:FNine}), SU(9)$'$ families appear for one ${\tsix}$ plus five $\nineb$'s,  after removing vector-like pairs, for which the SU(9)$'$ anomaly is $(+5-5\cdot 1)=0$ from  Eq. (\ref{eq:AnomSU}). 
Thus, the spectra in Eqs. (\ref{eq:FFive}) and (\ref{eq:FNine}) do not lead to non-Abelian gauge anomalies.  Lets us define the following U(1) generators,
\dis{
&Q_1 = \frac19\left((-2)^9;0,0;0^5\right)\\
&Q_2  = (0^9;-2,0;0^5 )\\
&Q_3 = (0^9;0,-2;0^5 )\\
&Q_X = \left(0^9;0,0;(-2)^5\right).\label{eq:u1Qs}
}
The eigenvalues of $Q_i$ for the state, for example $P+kV_0$, is calculated by
\dis{
\sum_aQ_i^a(P+kV_0)^a.
}

In Table \ref{tab:T7minus}, we illustrate a case explicitly how the chiralities and multiplicities are calculated, in the twisted sector $T_7^-$,
\dis{
T_7^-: \left( (\frac{+7}{12})^9;\frac{+5}{12}, \frac{+6}{12};(\frac{-2}{12})^5\right).
}
 For the spinor type SU(5) tensors,
\dis{
P_5=\left(-^9;--;\underline{+++--}\right):~\textrm{two indices tensor},\\[0.3em]
P_5=\left(-^9;--;\underline{++++-}\right):~\textrm{one index tensor},
}
where + and -- represent $+\frac12$ and $-\frac12$, respectively, and underline means all possible permutations among the entries, we have the non-singlet representations (\ref{eq:tenfive}) \cite{KimPRD21}.

 \begin{table}[!ht]
\begin{center}
\begin{tabular}{|cc|c|c|c|cc|c|c|c|}
 \hline &&& &&& &&&  \\[-1.15em]
  Chirality  &   $\tilde s$& $-\tilde{s}\cdot\phi_s$& $-p_{\rm vec}^{ k\,\rm th}\cdot \phi_s$ &$k\,P_{5}\cdot V_-$& $(k/2)\phi_s^2$,&$  -(k/2) V_-^2 $,& ~~$\Delta_{1}^{N},~-\delta_2^N
  $   &$\Theta_5,$& Mult. of SU(5)  \\[0.15em]
 \hline\hline &&& &&& &&&  \\[-1.15em]
$\ominus=L$& $(---)$  &  $\frac{+5}{12}$ & $\frac{+1}{12}$~&$\frac{-2}{12}$ &$\frac{+147}{144}  $  &$\frac{-651}{144};\frac{6}{12}$ & $ \frac{+2}{12},~~\frac{-2}{12}$&$\frac{+10}{12} $& $0\cdot\ten$\\ [0.1em]
 \hline &&& &&& &&& \\[-1.25em]
$\oplus=R$& $(+++)$  &  $\frac{-5}{12}$ & $\frac{+1}{12}$~&$\frac{-2}{12}$ &$\frac{+147}{144} $  &$\frac{-651}{144}$ & $ \frac{+2}{12},~~\frac{-2}{12}$&$\frac{0}{12}$&  $3\cdot\ten$  \\[0.15em] 
\hline\hline
$\ominus=L$& $(---)$  &  $\frac{+5}{12}$ & $~\frac{+1}{12}$~&~$\frac{-4}{12}$ &$\frac{+147}{144}  $  &$\frac{-651}{144}$ & $ \frac{+6}{12},~~\frac{-2}{12}$&$\frac{0}{12}$&  $3\cdot\five$\\ [0.1em]
 \hline &&& &&& &&& \\[-1.25em]
$\oplus=R$& $(+++)$  &  $\frac{-5}{12}$ & $~\frac{+1}{12}$~&~$\frac{-4}{12}$ &$\frac{+147}{144} $  &$\frac{-651}{144}$ & $ \frac{+6}{12},~~\frac{-2}{12}$&$\frac{-10}{12}$& $0\cdot\five$  \\[0.15em] 
\hline
\end{tabular}
\end{center}
\caption{Two indices (upper two rows) and one index (lower two rows)   spinor-forms from $V_7^-$, shown in Tables XIII and XIV of Ref. \cite{KimPRD21}. Counting of multiplicities is given in \cite{KimPRD21}.  } \label{tab:T7minus}
\end{table}

 To calculate phase needed for the calculation of multiplicities, columns are listed according to the order of terms in $\Theta_{\rm Group}$,
\begin{eqnarray}
&\Theta_{\rm Group}  =-\tilde{s}\cdot\phi_s -k\,p_{\rm vec}^{ k\,\rm th}\cdot \phi_{s} + k\,P\cdot V_-+\frac{k}{2}(\phi_s^2-V_-^2)+\Delta_k^N-  \delta_k^N ,\label{eq:PhaseA}
\end{eqnarray}
where 
\dis{
\phi_s&=\left(\frac{5}{12},\frac{4}{12},\frac{1}{12}  \right),\\[0.5em]
V_-&=\left(\frac{7}{12}, \frac{7}{12}, \frac{7}{12}, \frac{7}{12}, \frac{7}{12}, \frac{7}{12}, \frac{7}{12}, \frac{7}{12}, \frac{7}{12};\frac{5}{12}, \frac{6}{12};\frac{-2}{12}, \frac{-2}{12}, \frac{-2}{12}, \frac{-2}{12}, \frac{-2}{12} \right),\label{eq:Vsevmin}
 }
and
\begin{eqnarray}
\delta_k^N=2\delta_k .\label{eq:deltakN}
\end{eqnarray}
Here, $s$ is taken for the even number of $\pm$'s from $(\pm;\pm,\pm,\pm)$ where the first $\pm$ is $\oplus$ or $\ominus$ which determine the chirality as R and L, respectively. In the second row of  Table \ref{tab:T7minus}, for example, it is right-handed and $-s\cdot\phi_s=-(\frac12\cdot\frac{5}{12}+\frac12\cdot\frac{4}{12}+\frac12\cdot\frac{1}{12})=-\frac{5}{12}$. Other notations are given in detail in Refs. \cite{LNP954}, \cite{KimPRD21} and \cite{Kim20ijmpa}. For the SU(5) gauge group, the phase is given as $\frac{0}{12}$ that gives the multiplicity 3 in the twisted sector $T_7^-$.

 The spinor with the even number of +'s\footnote{The + and -   represent $\frac{+1}{2}$ and  $\frac{-1}{2}$, respectively.} for $\tenb$ and $\five$ are
\dis{
\tenb:&\left(-^9; -,-;\underline{+++--}\right) \\
{\five}:&\left(+^9; +,+;\underline{+----}\right). \label{eq:tenfive}
}
These satisfy $(P_5+6V_0)^2=\frac{216}{144} $  which saturates the masslessness condition, and in Eq. (\ref{eq:tenfive}) we show the entries for $\tenb$\,s and  $\five$\,s, which are listed again in Table \ref{tab:SMUones}. For the rest of spectra, we omit the details and just present the results in the tables.

\bigskip
\subsection{Searching for an acceptable vacuum suitable for the visible sector}

\begin{table}[t!]
\begin{center}
\begin{tabular}{@{}lc|ccc|c|cccc@{}} \toprule
 &State($P+kV_0$) &$\Theta_i$ & $(N^L)_j$&${\cal P}~~ {\bf R}_X$(Sect.)&$Q_R$ &$Q_1$&$Q_2$ &$Q_3$ &$Q_X$   \\[0.1em] \colrule
 $\tenb$  & $(-^9;-,-;\underline{+++--})$ &$\frac{+1}{12}$ &$3$ & $4~~{\tenb}_{-1}(T_7^-)_L$ &$+1$& $1$ & $1$ & $1$ & $-1$  \\
 $\tenb$  & $(-^9;-,-;\underline{+++--})$ &$\frac{-9}{12}$ &$3 $ & $5~~{\tenb}_{-1}(T_7^-)_R$ &$+1$& $1$ & $1$ & $1$ & $-1$   \\
 ${\five}$  & $(+^9;+,+;\underline{+----})$ &$\frac{+1}{12}$ &$3 $ & $4~~{\five}_{3}(T_7^-)_L$ &$+1$& $-1$ & $-1$ & $-1$ & $+3$ \\
 ${\five}$  & $(+^9;+,+;\underline{+----})$ &$\frac{-9}{12}$ &$3$ & $5~~{\five}_{3}(T_7^-)_R$ &$+1$& $-1$ & $-1$ & $-1$ & $+3$   \\
$\alpha$  & $(0^9)\left(-1,0;+1\,0^4\right)'$ &$\frac{+1}{12}$ &$3 $ & $4~~{{\five}_{h,3}}(T_1^-)_R$ &$-4$& $0$ & $+2$ & $0$ & $-2$\\
 $\beta$  & $(0^9)\left(+1,0;-1\,0^4\right)'$ &
 $\frac{-9}{12}$ &$3$ & $5~~{{\fiveb}_{h,-3}}(T_1^-)_R$ &$+4$& $0$&$-2$ & $0$ & $+2$ \\[0.2em]
 \hline
${\tsix}'$  & $(\underline{--+^7};+,-;-^5)$ &$0$ &$1(1_{\bar1}+1_3)$& $1~ {\tsix}'_{0}(T_{9}^0)_L$ &$-18 $& $-5$ & $-1$ & $+1$ & $+5$ \\
${\nine}'$  & $(\underline{-1\,0^8};0,0;(-1)^5)$ &$\frac{-6}{12}$  &$1(1_{\bar1}+1_3)$& $7~~ {\nine}'_{0}(T_{4}^0)_R$ &$+19$& $+2$ & $0$ & $0$ & $+10$\\
 ${\nine}'$  &$(\underline{+\,-^8};-,\frac{-3}2;-^5)$&$\frac{-9}{12}$  &$1(1_{\bar1}+1_3)$& $7~~ {\nine}'_{0}(T_{4}^-)_L$ &$+31$& $+7$ & $+1$ & $+3$ & $+5$ \\
 ${\nine}'$  &$(\underline{-1\,0^8};-1,0;0^5)$&$\frac{-6}{12}$  &$1(1_{\bar1}+1_3)$& $3~~ {\nine}'_{0}(T_1^-)_R$ &$ +7$& $+2$ & $+2$ & $0$ & $0$ \\${\nine}'$  &$(\underline{-1\,0^8};0,+1;0^5)$&$\frac{-5}{12}$  &$1(1_{\bar1}+1_3)$& $3~~ {\nine}'_{0}(T_2^0)_R$ &$+12$& $+2$ & $0$ & $-2$ & $0$ \\
${\nine}'$  &$(\underline{-\,+^8};-,-;-^5)$&$\frac{0}{12}$  &$1(1_{\bar1}+1_3)$& $3~~ {\nine}'_{0}(T_{7}^0)_R$ &$-29$& $-7$ & $+1$ & $+1$ & $+5$ \\
${\nine}'$  &$(\underline{-1\,0^8};+3,-4;0^5)$&$\frac{+6}{12}$  &$1(1_{\bar1}+1_3)$& $3~~ {\nine}'_{0}(T_{7}^+)_L$ &$+3$& $+2$ & $-6$ & $+8$ & $0$ \\
 ${\nine}'$  &$(\underline{-1\,0^8};0,1;0^5)$&$\frac{+5}{12}$  &$1(1_{\bar1}+1_3)$& $1~~~ {\nine}'_{0}(U)_L$ &$+12$& $+2$ & $0$ & $-2$ & $0$ \\[0.2em]
  \botrule
\end{tabular} 
\end{center}
\caption{U(1) charges of the non-singlet fields of Ref. \cite{KimPRD21}.  }\label{tab:SMUones} 
\end{table}
 
 As the Universe evolves, the vacuum with the least chiral fields is chosen. When the Higgs fields of the SM, $H_{u,d}$, obtain VEVs, the Yukawa couplings make the top and bottom quarks massive. In our supersymmetric case, these Yukawa couplings are possible with \UoR\,charge +2. Such  U(1)$_R$ of $Q_R=2$  for the visible sector fields  is satisfied with 
\dis{
Q_R=\frac92 Q_1-Q_2-\frac32 Q_3+Q_X.\label{eq:QR}
}
Non-singlet fields in the visible and hidden sectors are listed in Table \ref{tab:SMUones}.

Gauge symmetry allows the \flip tree level Yukawa couplings ${\tenb}_{-1}(T_7^-)_{R}{\tenb}_{-1}(T_7^-)_{R}{\fiveb}_{-3h}(T_1^-)_{R}$ with $Q_R=2$ modulo 4 for $b$-quark mass and $ {\tenb}_{-1}(T_7^-)_{R}{\five}_{3}(T_7^-)_R{\five}_{3h}(T_1^-)_{R}$ with $Q_R=2$ modulo 4 for $t$-quark mass are allowed.  For these, we must check the string selection rules given in Ref. \cite{KimPRD21}. However, in the $\Z_{12-I}$ orbifold compactification,
\dis{
{\tenb}_{-1R}(T_7^-){\tenb}_{-1R}(T_7^-){\fiveb}_{-3,hR}(T_1^-),~{\tenb}_{-1R}(T_7^-){\five}_{3R}(T_7^-){\five}_{3,hR}(T_7^-),\label{eq:YukRaw}
}
do not satisfy $\sum_i i(T_i)=0$ modulo 12. To satisfy the string selection rule, we multiply R-hand singlet(s) with $Q_R=0$ and $i(T_i)= 9$ and 3, respectively. The VEVs of these singlets should be of order the string scale. Note the chiral singlets listed in Table \ref{tab:ChiralSinglets}.  In Eq. (\ref{eq:YukRaw}), multiplying  $I^3$ to the first term and $C^3$ to the second term can satisfy the required conditions. Thus, $\langle I\rangle$ and $\langle C\rangle$
are at the string scale.

\begin{table}[t!]
\begin{center}
\begin{tabular}{@{}lc|ccc|c|cccc@{}} \toprule
 &State($P+kV_0$) &$\Theta_i$ &  &${\cal P}~~ {\bf R}_X$(Sect.)&$Q_R$ &$Q_1$&$Q_2$ &$Q_3$   &  $Q_{X}$ \\[0.1em] \colrule 
$A$  & $(+^9;+,-;-^5)$&$\frac{-4}{12}$ &  &$1~~{\one}_{0}(T_9^0)_L$  &$+50$& $-9$ & $-1$ & $+1$  &   $+5$  \\
$B$  & $(-^9;-,-;+^5)$&$\frac{+2}{12}$ & & $1~~ {\one}_{0}(T_9^+)_R$ &$+38$& $+9$ & $+1$ & $+1$ &    $-5 $  \\
$C$  & $(-^9;-,+;+^5)$&$0$ & &$1~~{\one}_{0}(T_9^-)_R$ &$+36$&$+9$ & $+1$ & $-1$ &   $-5$\\
$D$  & $(-^9;+,\pm;+^5)$&$0$ & &$6~~{\one}_{0}(T_4^+)_{L\oplus R}$  &$+38,+35$& $+9$ & $-1$ & $\mp 1$ & $  -5$\\
$E$  & $(-^9;+,\frac{-3~{\rm or}~-5}{2};-^5)$&$\frac{+1}{2}$ & & $14~~{\one}_{0}(T_4^-)_R $ &$+47,+39$& $+9$  & $-1$ & $+3~{\rm or}+5$ &   $+5$\\
$F$  & $(0^9;-1,-1;0^5 )$&$\frac{+1}{12}$ & &$3~~{\one}_{0}(T_1^-)_L $ &$-5$&$0 $ & $+2$  & $+2$ &  $0$\\
$G$  & $(0^9;+1,+1;0^5 )$&$\frac{+1}{12}$ & &$3~~{\one}_{0}(T_2^0)_R$  &$+5$& $0$ & $-2$ & $-2$  &    $0$ \\
$H$  & $(0^9;+1,+1;0^5 )$&$\frac{+1}{12}$ & & $3~~ {\one}_{0}(T_2^-)_R$ & $+5$ & $0$&$-2$   & $-2 $ &
$0$ \\
$I$  & $(+^9;\frac{-3}{2},-;-^5)$&$0$ & &$3~~{\one}_{0}(T_7^0)_L$ &$-40$&$-9$ & $+3$ & $+1$ &   $+5$\\
$J$  & $(-^9;+,-;+^5)$&$0$ & &$3~~{\one}_{0}(T_7^-)_L$  &$+45$& $+9$ & $-1$ & $+1$ &   $-5$\\
$K_1$  & $(+^9;\frac{-3}{2},\frac{-5}{2};+^5)$&$\frac{+1}{2}$ & & $21~~{\one}_{0}(T_6)_{L\oplus R} $ &$-56$& $-9$  & $+3$ & $+5$ &   $-5$\\
$K_2$  & $(+^9;\frac{-3}{2},\frac{-7}{2};+^5)$&$\frac{+1}{2}$ & &$21~~{\one}_{0}(T_6)_{L\oplus R} $ &$-59$&$-9$ & $+3$  & $+7$ &  $-5$\\
$K_3$  & $(+^9;\frac{-3}{2},\frac{-5}{2};-^5)$&$\frac{+1}{2}$ & &$21~~{\one}_{0}(T_6)_{L\oplus R}$  &$-61$& $-9$ & $+3$ & $+5$  &    $+5$ \\
$K_4$  & $(+^9;\frac{-3}{2},\frac{-7}{2};-^5)$&$\frac{+1}{2}$ & & $21~~ {\one}_{0}(T_6)_{L\oplus R}$ & $-51$ & $-9$&$+3$   & $+7$ &$+5$ \\
 ${\Sigma}_1$  & $(+^9;\frac{-3}{2},\frac{-5~{\rm or}-7}{2};\underline{++---})$& $\frac{+5}{12}$ &&$36~~ {\ten}_{+1}(T_6)_{L\oplus R}$ &$-49,-55$&$-9$ &  $+3,$ & $+3~{\rm or}~+7$ &  $+1$\\
 $\overline{\Sigma}_2$  & $(+^9;\frac{-3}{2},\frac{-5~{\rm or}-7}{2};\underline{+++--})$& $\frac{+5}{12}$ & &$36~~{\tenb}_{-1}(T_6)_{L\oplus R}$ &$-51,-57 $&$-9$   & $+3,$ & $+3~{\rm or}~+7$  & $-1$
  \\[0.2em]
  \botrule
\end{tabular} 
\end{center}
\caption{Summary of chiral singlets \cite{KimPRD21}. Neutral singlets   are singlets with $X=0$.  }\label{tab:ChiralSinglets} 
\end{table}

 \bigskip
\subsection{Hidden sector}\label{sec:HiddenS} 
The $Q_R$ values of $\nine'_R$ are
\dis{
\nine':~~&(T_4^0)_R,(T_1^-)_R, ,(T_2^0)_R, ,(T_7^0)_R,  \\
Q_R=& +19,~~+7,~~~~+12,~~~-29
}
When the confining force becomes strong, the SU(9)$'$ 
\begin{figure}[!h]\hskip -0.3cm
\includegraphics[height=0.35\textwidth]{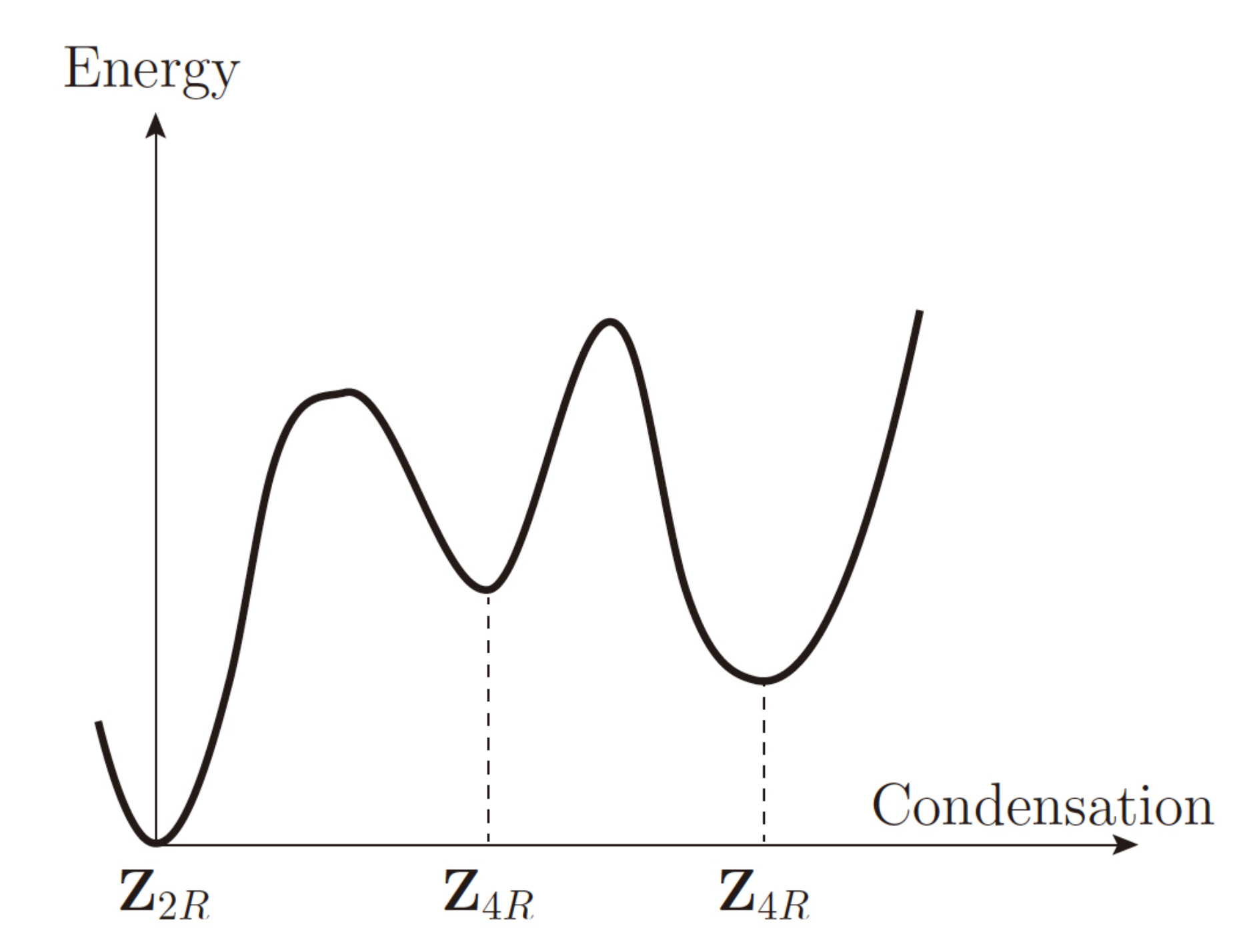}
\caption{A schematic view of discrete vacua at the condensation points.}\label{fig:Vacua} 
\end{figure}
condensates form and respect only the  SU(9)$'$ singlet condition. Since $Q_R(\tsixb'_R)=+18$, the $Q_R$ of following condensates are
\dis{
\tsixb'_R\nine'(T_4^0)_R \nine'(T_1^-)_R &:~+44 ~\textrm{modulo 12}=-4,\\
\tsixb'_R\nine'(T_4^0)_R\, \nine'(T_2^0)_R &:~+49 ~\textrm{modulo 12}=+1,\\
\tsixb'_R\nine'(T_4^0)_R\, \nine'(T_7^0)_R &:~+8 ~\textrm{modulo 12}=-4,\\
\tsixb'_R\nine'(T_1^-)_R \nine'(T_2^0)_R &:~+37 ~\textrm{modulo 12}=+1,\\
\tsixb'_R\nine'(T_1^-)_R \nine'(T_7^0)_R &:~-4\\
\tsixb'_R\nine'(T_2^0)_R\, \nine'(T_7^0)_R &:~+2.\\
}

Note that we do not require $\sum_i i(T_i)=0$ modulo 12 because the contents of effective fields are important at the confining scale which is far below the string scale.
The resulting $\Z_N$ are 1, 2 and 4. There are three vacua for the $\Z_{4R}$ symmetry. Supersymmetry requires the last case, possessing $Q_R=2$, as a possible superpotential. Therefore, if we choose the vacuum
\dis{
\Xi= \langle {\tsixb}'(T_9^0)_R\, {\nine}'(T_2^0)_R\, {\nine}'(T_7^0)_R\rangle, \label{eq:CondScale}
}
the potential energy is zero. At the minima of the other cases,  the potential energy are non-zero. It is shown schematically in Fig. \ref{fig:Vacua}. For simplicity, let us choose the followinmg as the surviving five $\nine'_R$'s,

Fundamental representations of SU(9)$'$ are   $7\cdot\nineb'(T_4^-)_R,3\cdot\nineb'(T_7^+)_R,\nineb'(U)_R,7\cdot\nine'(T_4^0)_R,3\cdot\nine'(T_1^-)_R,3\cdot\nine'(T_2^0)_R$, and $3\cdot\nine'(T_7^0)_R$. Eleven $\nineb'_R$'s combine to form vector-like representations with eleven linear combinations out of sixteen $\nine'_R$'s. Then, there remain five linear combinations of $\nine'_R$'s below the GUT scale.
\dis{
\nine'(T_4^0)_R,~ \nine'(T_1^-)_R ,~ \nine'(T_2^0)_R,~  \nine'(T_7^0)_R,~  \nine'(T_x)_R,~ \label{eq:fivenines}
}
where $x$ represents one linear combination.
Two of these five $\nine'_R$'s combine with $\tsixb'_R$ to form condensates. There are $_5C_2=10$ cases to form these condensates, for example,
\dis{
\langle \tsixb'_R(T_9^0)\cdot \nine'(T_i)_R\cdot  \nine'(T_j)_R\rangle = f^3 e^{i M_{ij}/f}\label{eq:Condsa36}
}
where $i,j$ represent the twisted sectors, $i,j=\{4,1,2,7,x \}$, and $f$ is of order the GUT scale. These ten mesons are exactly massless at this stage.
 For these ten mesons to obtain mass, the flavor symmetries of five  $ {\nine}'_R$'s of Eq. (\ref{eq:fivenines}
) should be broken explicitly. Explicit breaking terms are given above the GUT scale, where the breaking terms must satisfy the string selection rules. For the explicit breaking terms, we also use the VEVs of the SM Higgs fields, $\beta_R= H_d({\five}_{3h}, T_1^-)_R$ and  $\alpha_R= H_u(\fiveb_{-3h}, T_1^-)_R$.  The lowest order term we can consider from the string selection rule is
\dis{
W_{\rm ex-br}=\frac{1}{M}\,{\nineb}_0'(T_4^-)_R {\nine}_0'(T_2^0)_R {\one}_0(T_4^-)_R {\one}_0(T_2^-)_R . 
}
which, however, carries $Q_R=69$ or (61), and hence the above is not an allowed term at a supersymmetric vacuum.
An example is
\dis{
W_{\rm ex-br}=\frac{1}{M^4}\,{\nineb}_0'(T_4^-)_R {\nine}_0'(T_2^0)_R {\one}_0(T_4^-)_R ^4{\one}_0(T_2^-)_R , \label{eq:ExBreaking}
}
which carries $Q_R=-2$ with $Q_3= +5$ in Table  \ref{tab:ChiralSinglets}. If we replace the fields with the complex conjugated L-fields in (\ref{eq:ExBreaking}), then it is an allowed superpotential with $Q_R=2$. The VEV of  ${\one}_0(T_2^-)_R$ can be of order the GUT scale and its phase can be interpreted as a quintessential field by assigning an appropriate VEV to ${\one}_0(T_4^-)_R$. Out of many ${\one}_0$ singlets of Table \ref{tab:ChiralSinglets}, there are many possibilities for this kind of quintessential fields. 
  
Cosmological effects of quintessential axion is reviewed in
\cite{ChoiG20,Tsujikawa21}. From the effect of rolling-down  quintessential axion, the axion mass in the region $10^{-32}\,$eV--$10^{-20}\,$eV is excluded.
The interesting region of the  quintessential axion is marked as the inside of the dashed curve in Fig. \ref{fig:ULA}.

\section{Conclusion}\label{sec:Conclusion}
We  reviewed ideas of  quintessential axion as the dark energy source. We set a U(1) global symmetry from which a Goldstone boson is created by its spontaneous breaking. This Goldstone boson is made pseudo-Goldstone boson by explicit breaking terms of the  U(1) global symmetryinterpreted as a  quintessential axion if its decay constant is near the GUT scale and the energy density due to the explicit breaking term is O($(0.003\eV)^4$).  We tried to realize it in detail, obtaining SU(9)$'$ hidden sector gauge group for supersymmetry breaking, by a $\Z_{12-I}$ orbifold compactification of SO(32) heterotic string.

\acknowledgments{This work is supported in part by the National Research Foundation (NRF) grant  NRF-2018R1A2A3074631.}

   
\end{document}